\documentclass[5p,number]{elsarticle}
\usepackage{url}

\usepackage{amssymb}
\usepackage{amsmath}
\usepackage{float}
\usepackage{multirow}
\usepackage{makecell}

\journal{Acta Astronautica}

\begin{document}

\begin{frontmatter}



\title{Concept and Rationale for Stratospheric Balloon-Based Laser Debris Removal}


\author{F. Ruiz Vincueria $^{a,h}$} 
\author{M. Divoký $^{b}$} 
\author{E. Sánchez-Laulhé $^{a}$, H. Yang $^{a,h}$, J. Rodríguez $^{a,h}$} 
\author{V. Montero $^{c,d}$, V. Canales $^{c,d}$, E. Alcalde $^{c,d}$} 

\author{A. Zavoli $^{e}$} 
\author{M. Rico $^{f}$} 
\author{J. Klare $^{g}$} 
\author{A. Ollero $^{a}$} 

\affiliation{country={%
$^{a}$Asociación de Investigación y Cooperación Industrial de Andalucía (AICIA), Camino de los Descubrimientos s/n, Seville, 41092, Spain\\
$^{b}$HiLASE Centre, Institute of Physics of the Czech Academy of Sciences, Za Radnicí 828, Dolní Břežany, 252 41, Czech Republic\\
$^{c}$B2Space Launch Systems LTD, 19a Greenwich Rd, Newport, NP20 2NN, United Kingdom\\
$^{d}$B2Space Launch Systems SL, Calle Barrio S/N, Pineda de la Sierra, Burgos, 09199, Spain\\
$^{e}$Sapienza University of Rome, Piazzale Aldo Moro 5, Rome, 00185, Italy\\
$^{f}$Centro de Láseres Pulsados (CLPU), Calle del Adaja 8, Villamayor (Salamanca), 37185, Spain\\
$^{g}$Fraunhofer Institute for High Frequency Physics and Radar Techniques FHR, Fraunhoferstraße 20, Wachtberg, 53343, Germany\\
$^{h}$FRV Space Technologies SL, Plaza Vicente Aleixandre 8, Seville, 41013, Spain%
}}

\begin{abstract}
Laser-based deorbiting has long been proposed as a promising technique for mitigating the growing population of orbital debris, particularly decimetre-scale objects --- large enough to cause catastrophic collisions yet too small to be economically captured by current active debris removal (ADR) missions. However, the practical implementation of laser-based approaches remains constrained by several factors. Ground-based systems are affected by atmospheric absorption, scattering, and turbulence, which degrade beam quality and limit effective energy delivery. In contrast, space-based laser platforms offer significant advantages, as the improved propagation efficiency over long distances leads to more compact and lower-power systems. Nevertheless, they require complex, high-cost space infrastructures and the associated challenges in terms of deployment, maintenance, and operational scalability.

This paper proposes a stratospheric balloon--based concept for laser debris removal, exploring the intermediate operating regime between ground-based and space-based approaches. Operating above 99\% of the atmospheric mass, such a platform may enable ultraviolet (3HG/4HG) laser operation --- largely precluded from ground level by ozone absorption --- with improved propagation and ablation coupling efficiency, while presenting a different set of engineering constraints than orbital systems. Rather than presenting a fully developed system, this work focuses on the conceptual design and underlying rationale of the approach. We describe the main architectural elements, discuss the expected operational envelope, and position the concept with respect to existing ground-based and space-based solutions, identifying potential advantages as well as key technical challenges and limitations that require further investigation.
This work is developed in the context of STRATOLASER, an EIC Pathfinder Challenge project launched in 2025, which will run for four years and conduct a series of stratospheric experiments aimed at raising the technology to TRL 4.

\end{abstract}

\vspace{-5mm}



\end{frontmatter}



\section{Introduction}
\label{sec1}

The rapid growth of the orbital debris population has become a critical threat to the long-term sustainability of space operations. Decades of launches, fragmentation events, and accidental collisions have produced a congested environment in low Earth orbit (LEO), where more than 40,000 catalogued objects and millions of uncatalogued fragments remain a persistent hazard \cite{dhinakaran2025}. While large objects can be tracked and selectively removed, the debris population in the decimetre-scale presents a unique challenge: these fragments are capable of causing catastrophic collisions yet remain too small to be economically captured using current active debris removal (ADR) technologies \cite{svotina2023}.

A broad range of ADR techniques have been proposed, each addressing different debris regimes and operational requirements.

\begin{figure}[H]
\centering
\includegraphics[width=0.95\columnwidth]{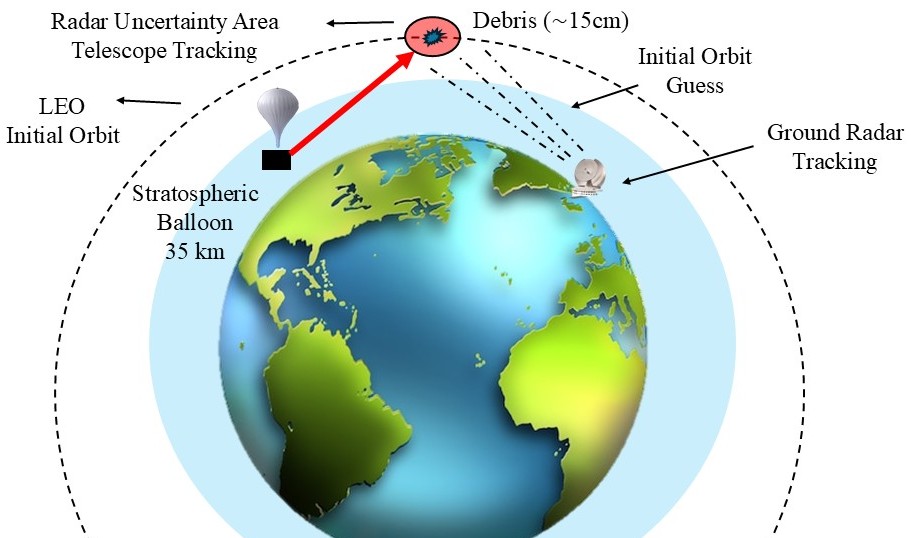}
\caption{Stratospheric laser ablation concept: a balloon at $\sim$35~km delivers pulsed UV laser energy to dm-scale LEO debris via ablation-induced momentum transfer. Operating above the ozone layer (which precludes UV transmission) enables near-diffraction-limited beam delivery and enhanced ablation coupling efficiency.}\label{fig1}
\end{figure}

\begin{figure*}[ht]
\centering
\includegraphics[width=1.75\columnwidth]{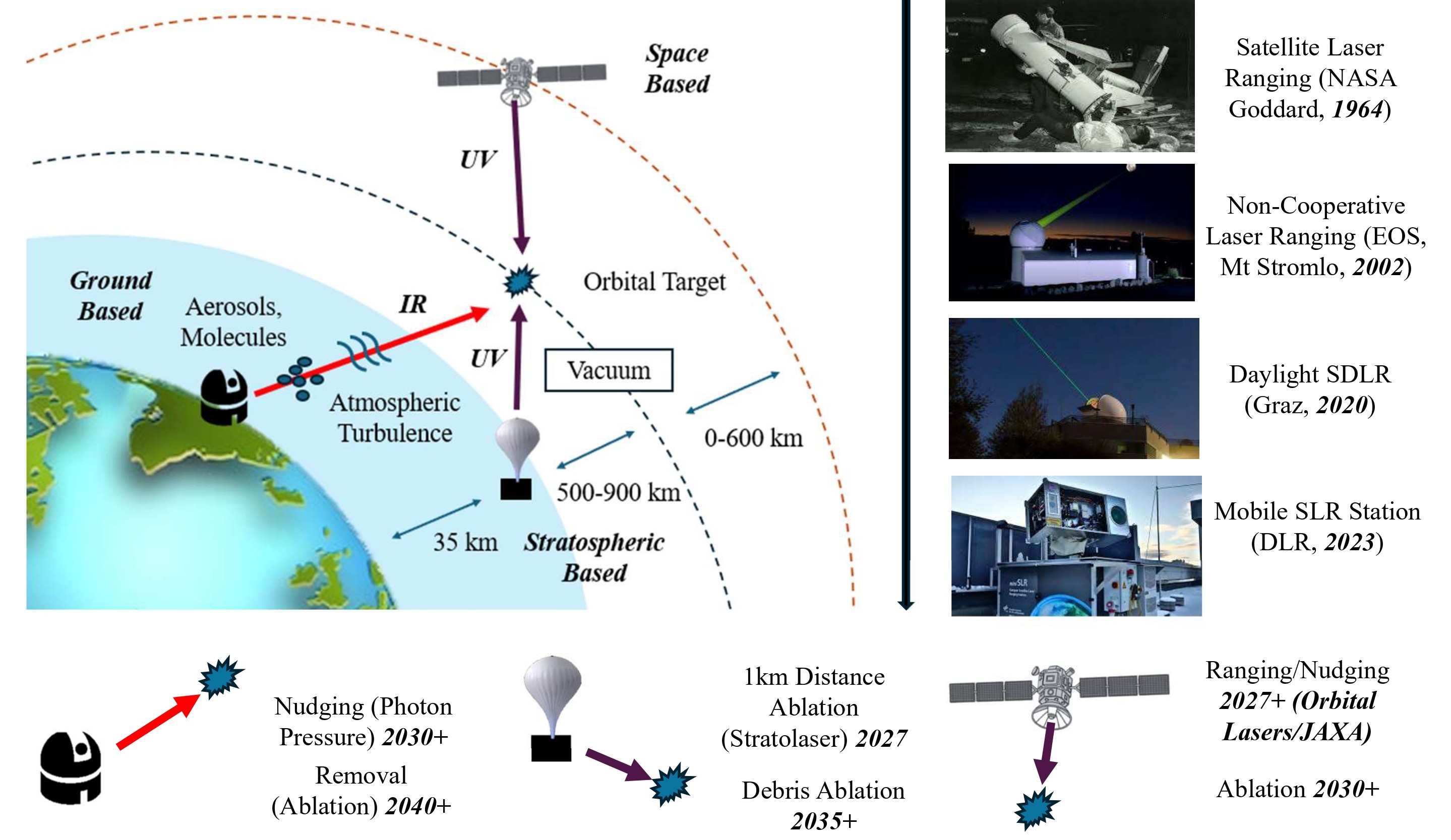}
\caption{Overview of laser-based orbital debris interaction from three platform geometries (ground-based, stratospheric, and space-based), illustrating atmospheric transmission constraints and engagement ranges. Also shown (right) are key milestones in non-cooperative space debris laser ranging --- the enabling first step toward momentum transfer operations --- from the first satellite laser ranging demonstration (NASA Goddard, 1964) to daylight SDLR (Graz, 2020) and the first cooperative mobile ranging station (DLR, 2024) \cite{steindorfer2020, eos, miniSLR2024}. Anticipated first demonstrations of laser momentum transfer are shown at bottom, including the STRATOLASER TRL4 ablation demonstration at 1~km distance (2027) \cite{stratolaser2025} and space-based ranging and ablation by Orbital Lasers/JAXA (2027+) \cite{jaxa}.}\label{fig2}
\end{figure*}

\begin{table*}[ht]
\centering
\setlength{\tabcolsep}{3pt}\begin{tabular}{llccccccc}
\hline
\textbf{Station} & \textbf{Country} & \textbf{First SDLR} & 
\textbf{$\lambda$} & \textbf{Laser} & \textbf{Pulse} & 
\textbf{Rate} & \textbf{Min. size} & \textbf{Precision} \\
\hline
AFRL/Starfire \cite{fugate1994} & USA & 1994$^*$ & --- & classif. & --- & --- & $\sim$m & --- \\
EOS Mt.~Stromlo \cite{greene2002} & Australia & 2002 & 1064~nm & 2.5~J & 5~ns & 100~Hz & 15~cm & cm--dm \\
Shanghai (SHAO) \cite{shao2013} & China & 2008 & 532~nm & $\sim$250~mJ & ns & 200~Hz & --- & 0.4--2.3~m \\
Graz \cite{steindorfer2020} & Austria & 2020 & 532~nm & 25~mJ & 10~ns & 1~kHz & $\sim$10~cm$^\dagger$ & $\sim$0.7~m \\
Iza\~{n}a-1 (ESA) \cite{izana1} & Spain & 2022$^\ddagger$ & IR & 150~mW & --- & --- & --- & --- \\
\hline
\end{tabular}
\caption{Representative laser-ranging stations and demonstrators relevant to non-cooperative SDLR 
capability. Laser parameters  correspond to the \emph{current} system 
configuration, not necessarily that of the first demonstration.
$^*$AFRL/Starfire (1994) demonstrated ranging of intact non-cooperative 
\emph{satellites} (metre-scale); technical parameters remain classified.
EOS Mt.~Stromlo (2002) constitutes the first publicly verified 
demonstration on centimetre-scale debris fragments \cite{greene2002}.
$^\dagger$Minimum size and precision entries for Graz are estimates under the cited operating conditions.
$^\ddagger$ESA's Iza\~{n}a-1 station was accepted in 2022 and currently operates at 150~mW for retroreflector satellites; 
ESA describes a planned 50~W upgrade for non-cooperative debris tracking, but no public minimum-size or precision value is reported.
}
\label{tab:sdlr}
\end{table*}

Satellite-based capture systems—using nets, harpoons, or robotic arms—have been demonstrated in missions such as RemoveDebris \cite{aglietti2020} and ELSA-d \cite{forshaw2019amos}.
Meanwhile, the upcoming ClearSpace-1 mission aims to be the first to capture and deorbit a large non-cooperative object \cite{vasconcelos2025}. While these methods provide controlled removal of selected large debris, they remain expensive and are not scalable to the thousands of medium-size fragments present in LEO.

Other ADR concepts include the Ion Beam Shepherd (IBS) \cite{bombardelli2011}, which imparts non-contact thrust using a quasi-neutral plasma beam, and electrodynamic tether systems \cite{sanmartin2010}, which exploit the interaction with Earth’s magnetic field to generate drag. Solar-sail concepts have also been proposed as a passive long-term solution for end-of-life deorbiting \cite{lappas2011}, though they require long timescales and are constrained by orbital geometry. Despite progress across these methods, none provide an operationally or economically scalable solution for the high-risk decimeter-range debris population.

Laser-based approaches, particularly pulsed-laser ablation, constitute a promising alternative. By directing high-energy pulses at the debris surface, material is vaporized and expelled, generating small momentum impulses that can accumulate to produce measurable orbital changes \cite{phipps2011}. Ground-based laser concepts have shown potential but suffer from long-standing physical constraints such as atmospheric absorption, turbulence, and pointing challenges \cite{tuer1982}. Space-based laser platforms can be used for re-entering or nudging debris \cite{phipps2016} with excellent propagation conditions using solar power. The main challenges lie in the cost and complexity of deploying and maintaining such infrastructure in orbit, as well as in the substantial power generation required from the solar arrays and the dissipation of the heat produced by the laser in vacuum. In this context, dedicated ad hoc laser architectures have been proposed, such as those based on coherent beam combined architectures, which have the potential to significantly improve the system capabilities (ICAN) \cite{soulard2014}. 

Stratospheric platforms --- operating at 30--40~km altitude (as presented in Figure~\ref{fig1}), above 99\% of the atmospheric mass --- represent a largely unexplored intermediate solution space. The closest precedent in the literature is the relay mirror concept proposed by Martin et al.\ \cite{martin2009}, in which a passive relay mirror mounted on a high-altitude aerostat redirects a ground-based laser beam from above the densest atmospheric layers; however, that architecture retains the ground-based laser source operating in the infrared, as UV wavelengths cannot propagate through the lower atmosphere without prohibitive losses, and the relay mirror still requires a 3--5~m aperture to deliver sufficient fluence at debris ranges --- leaving the fundamental energy-per-pulse requirement unchanged. A fully autonomous stratospheric laser platform --- in which the laser source itself (or at least the conversion stage) is elevated above the absorbing atmospheric layers --- enables operation at UV wavelengths, where ablation coupling efficiency is highest \cite{phipps2014}. At these altitudes, optical transmission is dramatically improved \cite{zhang2024}, platform recovery, maintenance, and reuse do not carry the operational burden of orbital maintenance, and the combination has the potential to enable high-cadence momentum transfer operations supporting the deorbiting of small objects in low Earth orbit, as well as complementary applications such as laser ranging and nudging.

In this paper, we present the concept and rationale for stratospheric balloon-based laser debris removal as a high-level conceptual description of the system, including its expected performance envelope and a comparative techno-economic analysis against ground-based and space-based laser approaches. The concept is being developed within STRATOLASER, an EIC Pathfinder Challenge project (``Strengthening the sustainability and resilience of EU space infrastructure'') launched in 2025, which will carry out a four-year programme of stratospheric balloon experiments aimed at maturing the technology to TRL~4 \cite{stratolaser2025}.

\section{State of the Art in Laser-Based Debris Removal}

The concept of using high-power lasers to impart momentum to orbital debris has been explored for more than three decades, with significant progress in modelling, experimental demonstrations, and system-level analyses. The fundamental mechanism relies on laser ablation, where short (nanosecond regime), intense pulses irradiate the surface of the debris, vaporizing a thin layer of material and generating a plume of ejecta that produces a recoil impulse. By repeating these impulses over multiple passes, the perigee of the debris can be progressively lowered, leading to atmospheric reentry and disposal.

One of the earliest operationally oriented concepts for laser-based debris removal was proposed by Phipps \cite{phipps1996}, who introduced the use of a 20-kW, 530-nm repetitively pulsed laser to induce momentum transfer via surface ablation. Subsequent work at Photonic Associates LLC over more than two decades established the underlying physics \cite{phipps2022a}, including impulse coupling and ablation regimes \cite{phipps2000, phipps2007}, with a focus on laser propulsion for space. More recent studies have addressed system-level architectures, including both ground- and space-based implementations and the use of ultraviolet lasers to improve efficiency \cite{phipps2012, phipps2013, phipps2014, phipps2016}. Recent work has further extended this framework to space debris traffic management applications, including laser ranging and collision avoidance nudging \cite{phipps2022b}.

The French space agency CNES has been a central institutional contributor to the operationalization of laser-based debris removal concepts. Bonnal et al.\ \cite{bonnal2013} established a systematic framework of active debris removal requirements, against which laser-based approaches were subsequently evaluated. Collaborative work between Bonnal and Photonic Associates LLC produced a spaceborne pulsed UV laser architecture for re-entering small LEO debris and nudging larger objects \cite{phipps2016}. Subsequent work extended this framework to space traffic management, proposing laser ranging and just-in-time nudging as operational tools for conjunction avoidance \cite{bonnal2020jtca, bonnal2020stm, phipps2022b}.

The German Aerospace Center (DLR) has contributed systematically to the physics of laser-based debris removal through the work of Scharring, Lorbeer, and Eckel. Their research has addressed impulse predictability for irregularly shaped targets \cite{scharring2016}, thermal constraints arising from heat accumulation under repetitive pulsing \cite{scharring2018}, and experimental validation of ablative momentum transfer at high energy levels \cite{lorbeer2018, scharring2019}. More recent work has examined overall system feasibility and the influence of material and surface properties on thermo-mechanical coupling \cite{scharring2023, scharring2024}, consolidating DLR as a key reference for the engineering characterization of ground-based laser removal concepts.

Other contributors have worked on laser--matter interaction for space debris materials along two complementary lines: material-specific numerical and experimental studies of coupling coefficients, and large-scale high-energy experimental campaigns. On the research side, Jin et al.\ \cite{jin2013} investigated ablation thresholds and impulse coupling for spacecraft materials including aluminum, titanium, and carbon fiber composites, confirming the existence of an optimal fluence regime consistent with Phipps' scaling law. This was extended by a multi-institutional campaign at the LULI facility (Ecole Polytechnique), where impulse coupling coefficients were measured for multiple materials at femtosecond and picosecond pulse durations, resolving significant discrepancies in previously reported values \cite{phipps2017luli}.

On the experimental campaign side, Le Bras et al.\ \cite{lebras2024} conducted ballistic pendulum experiments in vacuum on a broad set of metallic and carbon targets combined with hydrodynamic simulations, providing a systematic dataset of coupling efficiency as a function of laser parameters and target material. More recently, an international collaboration led by Boyer (CNRS) conducted an open-access experimental campaign on the HiLASE BIVOJ beamline (100~J, 10~ns, 1030~nm), investigating optimal thrust generation for operationally representative debris materials under repetitive pulsed irradiation \cite{boyer2024}.

Beyond system-level architectures, several authors have contributed specific modeling advances for space-based laser concepts. Soulard et al.\ \cite{soulard2014} proposed the use of a coherent fiber-laser array based on the ICAN architecture, showing that its high repetition rate and compactness could enable single-encounter deorbit of centimeter-scale debris from a spaceborne platform. Schmitz et al.\ \cite{schmitz2015} developed a parametric mission performance model for a space-based debris sweeper using an Nd:YAG laser, identifying laser range distance and target catalog availability as principal drivers of concept viability. Gambi and García del Pino \cite{gambi2017} addressed the guidance problem, deriving post-Newtonian equations of motion for space-based acquisition, pointing, and tracking systems and quantifying the pointing accuracy degradation introduced by lateral velocity components of the target relative to the laser platform. More recently, Walker and Vasile \cite{walker2023} investigated the combined use of photon pressure and laser ablation from an opportunistic space-based concept requiring no prior knowledge of debris orbits, simulating a 10-year mission lifetime and analyzing the effects of off-axis momentum transfer on debris attitude.

\begin{figure}[t]
\centering
\includegraphics[width=0.9\columnwidth]{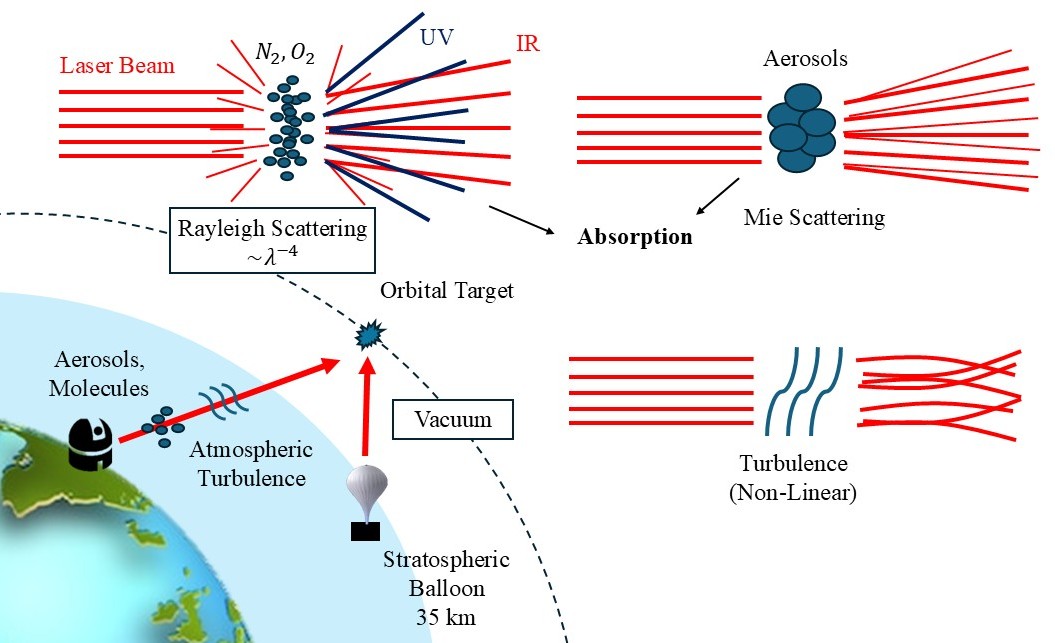}
\caption{\textbf{Atmospheric effects degrading laser propagation from the ground.} 
    Rayleigh scattering by air molecules ($\propto \lambda^{-4}$), Mie scattering by aerosols, absorption, 
    and nonlinear turbulence distort or attenuate the beam. 
    A stratospheric platform at $\sim$35 km avoids the lower atmosphere, enabling propagation 
    through much lower-density air and preserving beam quality.}\label{fig3}
\vspace{-2mm}
\end{figure}

Hybrid ground--space architectures have also been proposed as a way to exploit the complementary strengths of both platforms. Wen et al.\ \cite{wen2017} combined a ground-based and a space-based laser in a coordinated removal scheme, establishing an impulse coupling model that accounts for debris rotation and deriving an orbital momentum transfer model to optimize the deorbit sequence. Numerical simulations demonstrated that small-scale debris at 800 km altitude could be removed within a single pass using approximately 1500 laser pulses, suggesting that hybrid coordination can outperform either platform operating independently.

A related hybrid concept was proposed by Martin et al.\ \cite{martin2009} at Boeing, 
in which a passive relay mirror mounted on a high-altitude aerostat at $\sim$32~km 
redirects a ground-based infrared laser beam from above the densest atmospheric layers 
toward LEO debris. This architecture reduces the adaptive optics requirements on the 
ground system and extends the effective engagement range; however, it retains the 
ground-based infrared laser source --- as UV wavelengths cannot propagate through the 
lower atmosphere without prohibitive losses --- and the relay mirror still requires a 
3--5~m aperture to deliver sufficient fluence at debris ranges, leaving the fundamental 
energy-per-pulse requirement unchanged.

\subsection{Funded Projects and Demonstrators}

The studies reviewed above highlight both the potential and the engineering 
challenges of laser-based debris removal, motivating a number of funded 
projects and demonstrators (see Figure~\ref{fig2}). The CleanSpace Project 
\cite{cleanspace}, led by CILAS (France) and funded by the EU FP7 (2011--2014), 
investigated ground-based high-energy laser stations for momentum transfer on 
1--10~cm LEO debris. On the space-based side, Orbital Lasers (a SKY Perfect 
JSAT--RIKEN spinout under the JAXA J-SPARC framework) is developing a prototype 
in-orbit laser-ablative demonstrator targeting a 2027 launch and commercial 
Active Debris Removal services by 2029 \cite{jaxa}. Separately, ESA is 
consolidating laser ranging and momentum-transfer activities under the OMLET 
(Orbital Maintenance via Laser momEntum Transfer) project, which combines 
space debris laser ranging with a roadmap toward laser-based collision avoidance 
services for satellite operators \cite{omlet}. Finally, the present STRATOLASER 
project will perform the first demonstration of laser ablation from a 
stratospheric platform, advancing the balloon-based concept to TRL~4 
\cite{stratolaser2025}.

\subsection{Laser Ranging of Space Debris: The Enabling First Step}

Precise orbit determination of non-cooperative debris via space debris laser 
ranging (SDLR) is the first step towards any planned laser ablation or nudging 
intervention. Cooperative satellite laser 
ranging (SLR) --- in which short laser pulses are timed against retroreflectors 
on designated satellites --- has been performed routinely since 1964, achieving 
millimetre-level precision \cite{steindorfer2025}. SDLR extends this principle 
to non-cooperative targets, detecting diffuse reflections that are typically six 
to eight orders of magnitude weaker than retroreflector returns \cite{zhang2024}. This fundamental 
difference drives a distinct engineering regime: whereas cooperative SLR operates 
at pulse energies of order 0.1~mJ with picosecond pulses, SDLR of debris 
objects requires energies in the 100~mJ--2.5~J range  to 
compensate for the weak diffuse return. Alternatively, systems operating at 
picosecond pulse durations with kHz--MHz repetition rates and single-photon 
detectors can achieve sub-decimeter precision at lower pulse energies, exploiting 
the high measurement rate.

The first public demonstration of non-cooperative SDLR was reported by Greene 
(EOS Australia) in 2002, achieving ranges to 15~cm objects at 1250~km using a 
76~cm telescope \cite{eos}. First demonstrations in China followed at Shanghai 
Astronomical Observatory in 2008, and in Europe at Graz Observatory in 2011. 
Daylight SDLR --- previously restricted to the narrow twilight window --- was 
first demonstrated at Graz in 2020 using real-time sky background correction 
\cite{steindorfer2020}. Most recently, a megahertz bistatic configuration at 
Graz demonstrated combined cooperative SLR and non-cooperative SDLR in a single 
setup, dramatically increasing return rates and operational flexibility 
\cite{steindorfer2025}.

Despite this progress, fewer than ten facilities worldwide have 
demonstrated non-cooperative SDLR capability. Table~\ref{tab:sdlr} summarises the key parameters of the principal stations 
with demonstrated SDLR capability. The contrast between nanosecond high-energy 
systems (EOS, Graz SDLR) and picosecond low-energy systems (Shanghai) reflects 
the fundamental trade-off between pulse energy and timing precision: nanosecond 
pulses require higher energy but are limited to $\sim$0.7~m ranging precision, 
while picosecond systems achieve sub-decimetre precision at lower energy through 
statistical averaging of high-repetition returns.

\begin{figure*}[t]
\centering
\includegraphics[width=2\columnwidth]{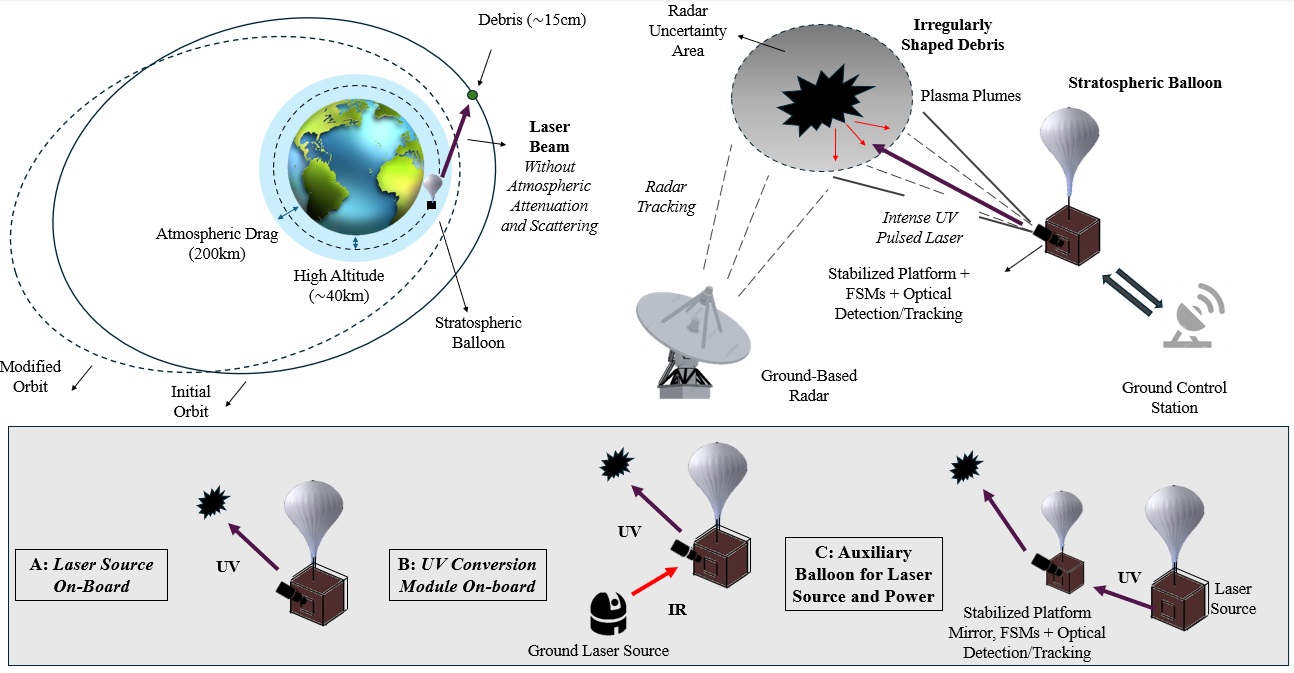}
\caption{Stratospheric laser ablation concept for orbital debris momentum transfer. A high-altitude balloon platform ($\sim 35-40~km$) delivers UV laser pulses to dm-scale LEO debris, operating above nearly all atmospheric attenuation and ozone absorption. Ground radar provides initial orbit estimates, refined onboard via telescope tracking and fast-steering mirrors prior to engagement. Three payload architectures are considered (bottom): (A) laser source fully on-board; (B) ground-based infrared laser source with on-board UV frequency conversion, extending the Boeing relay mirror concept \cite{martin2009} to an autonomous stratospheric platform; and (C) distributed configuration with laser source and beam director on separate balloons.}
\label{fig4}
\vspace{-2mm}
\end{figure*}

The only publicly known mobile laser ranging system, the DLR miniSLR \cite{miniSLR2024} --- a 600~kg transportable DPSSL-based station developed at the DLR Institute of Technical Physics --- is currently designed for cooperative SLR operation; extension to non-cooperative SDLR has not yet been demonstrated, leaving mobile non-cooperative ranging an open capability gap.

\section{Problem Statement}

\subsection{Limitations of Ground-Based Laser Systems}

Ground-based lasers operate through the densest layers of the 
atmosphere, where absorption, scattering, and turbulence severely 
degrade beam quality and reduce the fluence delivered to LEO targets. 
Figure~\ref{fig3} illustrates the main atmospheric effects: Rayleigh 
scattering, caused by air molecules such as N\textsubscript{2} and 
O\textsubscript{2}, whose cross-section scales as $\lambda^{-4}$ and 
increases rapidly toward shorter wavelengths; molecular absorption, 
most critically ozone (O\textsubscript{3}), which introduces deep 
absorption bands in the UV that render wavelengths below $\sim$320~nm 
effectively unusable from ground level, while H\textsubscript{2}O and 
CO\textsubscript{2} impose additional constraints in the infrared; and 
Mie scattering from aerosols and particulates, which remains significant 
across all wavelengths. Atmospheric turbulence further causes rapid phase 
distortions across the wavefront that fundamentally limit beam quality; 
although adaptive optics systems can partially compensate these effects, 
correcting a beam over a large aperture requires thousands of actuators 
\cite{phipps2011}, and residual errors remain significant at the pulse 
energies required for ablation. These processes collectively force 
ground-based systems to operate in the near-infrared, where atmospheric 
transmission is more favourable, at the cost of substantially reduced 
ablation coupling efficiency, as discussed later; Figure~\ref{fig5} quantifies the altitude dependence of atmospheric transmittance for the relevant wavelengths.~

Even at high-altitude observatories such as the ESA Iza\~{n}a station 
on Tenerife or Mount~Stromlo, acceptable transmittance is only achievable 
near zenith, during favourable meteorological conditions. To deliver sufficient fluence at LEO altitudes, the 
required pulse energy scales as $E \propto \Phi_{\mathrm{th}} 
\left(\lambda R\right)^2 / D^2$ \cite{phipps2011}, where 
$\Phi_{\mathrm{th}}$ is the ablation threshold fluence, $R$ the slant 
range, and $D$ the aperture diameter. The reference ground-based ablation 
design of Phipps et al.\ requires a 13-m aperture and multi-kilojoule 
pulses at ${\sim}10$~Hz \cite{phipps2011}; reducing the aperture would lead to higher energy requirements. At the same time, each debris object 
is accessible only during brief zenith passes of a few minutes per day, 
and ground-based systems remain heavily dependent on weather conditions, 
imposing further constraints on operational cadence.

In the context of laser ranging, ground-based systems represent today 
the most mature and operationally advanced approach, as evidenced by the 
growing network of SDLR stations described in Table 1, and remain the most likely near-to-medium term pathway for orbit 
determination of non-cooperative debris, provided that remaining 
limitations in coverage, daylight operation, and precision are 
progressively addressed \cite{steindorfer2025}.
For laser ablation, the outlook is considerably more challenging. 
Due to the mentioned constraints,  Scharring et al.\ \cite{scharring2023} show that even a 
nine-station global network operating near current technological 
limits could remove only $\sim$1,900 fragments per year, significantly below the rate estimated to stabilise the LEO 
environment. While nudging via photon pressure, which requires orders of 
magnitude less fluence, may become feasible from ground earlier; full ablative removal, however, remains a 
longer-term prospect, motivating the exploration of alternative 
platform geometries with shorter engagement ranges and improved 
transmission.

\subsection{Space-Based Laser Systems}

Space-based laser architectures offer compelling advantages over ground-based systems: they largely avoid lower-atmosphere transmission losses, benefit from excellent beam propagation efficiency, and can operate at UV wavelengths — where ablation efficiency is highest — without spectral or transmission penalties. Operating at ranges of 250~km for small debris and up to 600~km 
for large objects, systems such as the L'ADROIT concept \cite{phipps2014} 
demonstrate that pulse energies of 380~J to 2~kJ at 355~nm with a 
1.5~m aperture are sufficient to reach the ablation threshold of 
typical debris materials, representing a reduction of several orders of magnitude relative to ground-based requirements (specially in terms of mirror diameter). A complementary space-based design by Phipps and Bonnal \cite{phipps2016} extends this concept to a dual-mode system capable of nudging multi-ton derelict objects in LEO to avoid predicted conjunctions and raising defunct GEO satellites to graveyard orbits, operating with 8--10~kW average optical power from a solar-powered UV laser — at an estimated system cost of order \$200~M. A further architecture, proposed by Ebisuzaki et al.\ \cite{ebisuzaki2015}, combines the wide-field EUSO telescope with a scalable coherent fibre-laser array (CAN) initially deployed on the ISS, and progressively scaled to a dedicated free-flyer in polar orbit at $\sim$800~km capable of engaging debris at ranges up to 100~km. Furthermore, a platform in sun-synchronous orbit naturally accesses the full population of LEO debris without requiring orbital plane changes, enabling systematic engagement of the debris field over time \cite{schmitz2015}.

Nevertheless, space-based laser systems face significant engineering challenges. Power generation and thermal dissipation in orbit are particularly constraining: achieving average optical outputs around 15~kW at high repetition rate (example of L'ADROIT \cite{phipps2014}), demands electrical input power of at least 48~kW, even if it can be reduced with larger mirrors \cite{phipps2016}. In vacuum, heat can only be rejected via thermal radiation, and at typical spacecraft operating temperatures ($\sim350~K$) the Stefan--Boltzmann law yields a rejection capacity of approximately $350~W/m^2$ — implying that dissipating even 20~kW of waste heat requires radiator areas of at least 60~$m^2$ or more, with a corresponding mass and structural impact on the spacecraft. As a reference, the ISS rejects $\sim$70~kW continuously through several hundred square metres of radiator panels. In any case, advances in high-temperature radiators or Liquid Droplet Radiators will be key in determining the sizing of the final system.

Maintenance and repair are also substantially more difficult than for ground-based infrastructure, though the emergence of on-orbit servicing missions — such as Astroscale's ADRAS-J, Northrop Grumman's MEV, and ESA's ClearSpace-1 — suggests that this constraint may be progressively relaxed in the future. Finally, as shown in the analysis by Gambi and García del Pino \cite{gambi2017}, the relative angular rate between the laser platform and nearby debris objects imposes stringent requirements on the acquisition, pointing, and tracking system, and the first-order relation $\dot{\theta}=v_{\perp}/R$ shows that the burden is set by range and transverse relative velocity, not by the platform label alone. The shorter orbital standoff distance can therefore increase rates, while the same physics applies to ground and stratospheric platforms. velocity components are significant — a constraint that merits careful consideration in mission design.

\subsection{The Ground--Space Gap}

In summary, ground-based laser systems benefit from ease of maintenance, established infrastructure, and the possibility of incremental upgrades, but their effectiveness is fundamentally limited by atmospheric beam degradation, restricted twilight observation windows, and the large apertures and pulse energies required to compensate for propagation losses \cite{phipps2011, scharring2023}. Space-based systems largely avoid these lower-atmosphere propagation constraints and can operate at UV wavelengths with compact optics, but introduce a distinct set of engineering challenges. Both architectures present well-understood trade-offs, and active R\&D programmes are underway for each; yet the specific combination of near-space beam quality, operational flexibility, and low recurring cost required for scalable dm-scale debris mitigation remains an open challenge that motivates the exploration of intermediate platform geometries.

Stratospheric platforms at 30--40~km altitude offer a physically distinct operating regime that partially addresses the limitations of both ground- and space-based architectures. By operating above 99\% of the atmospheric mass, they substantially reduce lower-atmosphere turbulence and scintillation, easing (but not eliminating) the need for adaptive-optics correction, and enable much higher beam transmission at UV wavelengths --- where ground-based operation is precluded by ozone absorption but where ablation coupling efficiency is highest. The residual atmosphere, while insufficient for convective cooling of high-power components, provides a thermal environment less constraining than the hard vacuum faced by space-based systems, where waste heat rejection relies exclusively on thermal radiation. Moreover, unlike satellite platforms, they remain fully recoverable, re-launchable, and maintainable, allowing iterative hardware upgrades at a fraction of satellite development cost. The installation of optical components in the stratosphere has been done in the past. Beyond astronomy telescopes, experimental campaigns for optical detection of 
resident space objects from stratospheric balloons have been carried out \cite{kunalakantha2023, chianelli2024}, validating the near-space environment as a 
viable vantage point for space situational awareness.

Unlike ground-based stations, stratospheric platforms benefit from a significantly darker sky background, substantially relaxing the twilight constraint that limits ground-based SDLR operations and enabling more continuous operational availability for both ranging and ablation applications. They can be positioned over regions lacking fixed ground infrastructure, increasing network redundancy and improving the geometric diversity of ranging measurements required for high-precision orbit determination. Stratospheric winds impose constraints on geographic positioning that must be accounted for in mission planning; however, since Earth's rotation provides access to all orbital inclinations below the platform latitude regardless of east-west drift, wind-driven trajectory evolution can be incorporated into operational planning rather than treated as a fundamental limitation. From an engagement geometry perspective, a stratospheric platform is effectively stationary relative to the debris --- its lateral velocity of order 10--30~m/s being negligible compared to the orbital velocity of $\sim$7.5~km/s. This is fundamentally different from space-based operation, where the platform-debris relative velocity (and smaller distance)  limits the accessible engagement window \cite{gambi2017}. The stratospheric case is geometrically closer to ground-based operation, with manageable angular tracking rates throughout the approach phase.

These advantages must be weighed against constraints specific to stratospheric operation. Current large super-pressure balloons support payloads of approximately 3,000 ~kg \cite{kunalakantha2023}, which places demanding constraints on the mass, volume, and available electrical power of the laser system and its associated optics, thermal management, and pointing subsystems. However, the low recurring cost of stratospheric platforms means that distributing the system across two or more coordinated balloons --- separating, for instance, the laser source from the beam director, or the ranging station from the nudging payload --- represents a potentially viable path to scaling performance beyond what a single platform can support.

A further constraint is that for typical LEO debris populations at 600--1000~km altitude, engagement from a stratospheric platform at low elevation angles implies slant ranges of several thousand kilometres, driving laser energy and aperture requirements to impractical levels. Effective operation is therefore confined to a limited angular window around zenith, which reduces flexibility to optimize the impulses provided to the debris. 

\begin{figure}[t]
\centering
\includegraphics[width=0.95\columnwidth]{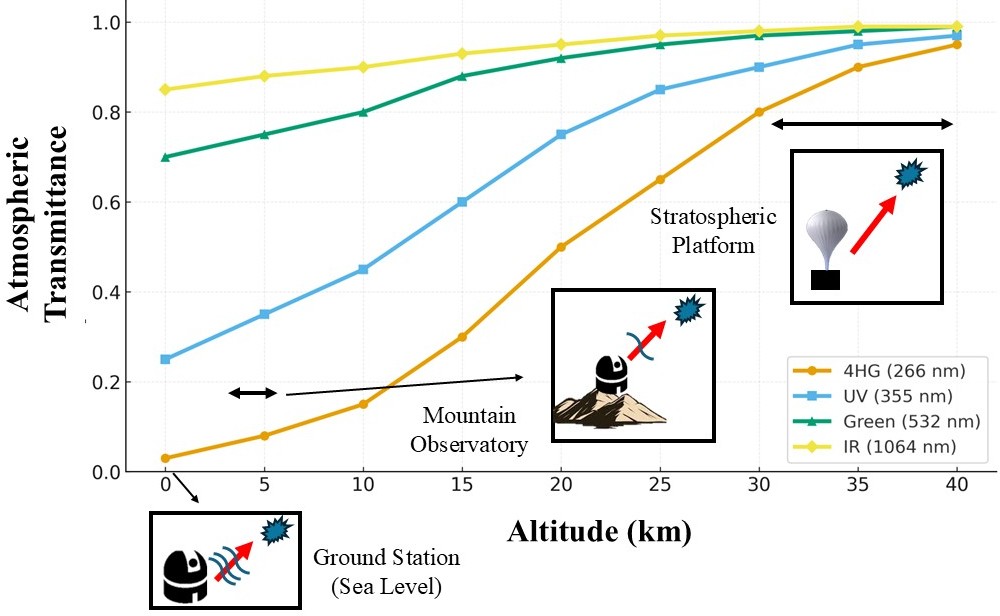}
\caption{\textbf{Atmospheric transmittance versus altitude for representative laser wavelengths.} 
    Shorter wavelengths such as 4HG (266 nm) and UV (355 nm) suffer strong attenuation at SL, 
    while transmittance increases sharply with altitude and approaches unity in the stratosphere ($\sim$35 km). 
    This enables efficient UV/DUV propagation compared with ground-based systems.}\label{fig5}
\vspace{-2mm}
\end{figure}

\begin{figure}[t]
\centering
\includegraphics[width=0.95\columnwidth]{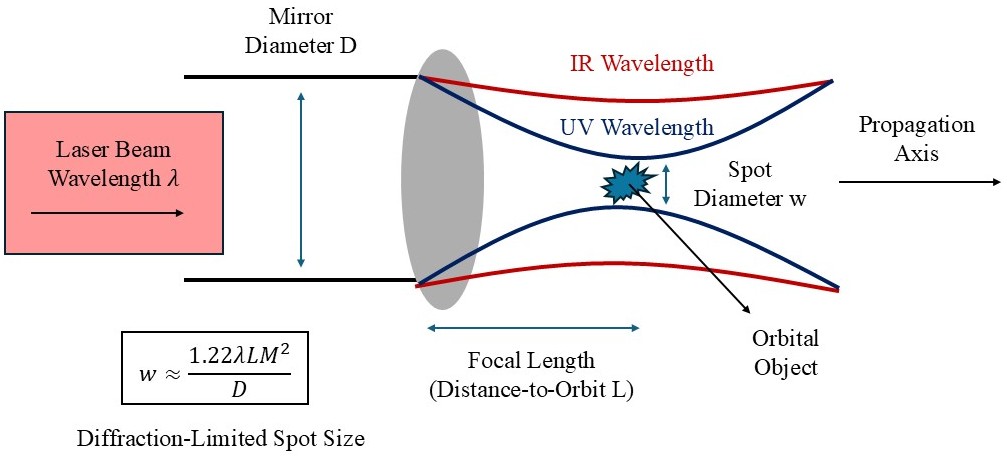}
\caption{\textbf{Diffraction-limited focusing performance for IR and UV wavelengths.}
    For a given mirror diameter $D$ and propagation distance $L$ (orbit altitude),
    shorter wavelengths achieve significantly smaller focal spot sizes,
    increasing the on-target fluence and thereby enhancing ablation efficiency
    for momentum transfer to orbital objects.
    The trend shown uses the first-null (Airy) convention
    $w \approx 1.22\,M^2\lambda L/D$; the quantitative spot radii and coupling
    fractions in Table~\ref{tab:spot} use the embedded-Gaussian form of
    Eq.~(\ref{eq:spot}), which is smaller by a factor $\approx 1.9$.}\label{fig6}
\vspace{-2mm}
\end{figure}

\section{Conceptual Design}

\subsection{Platform and Mission Architecture}

The proposed system (Figure~\ref{fig4}) is based on a long-duration 
stratospheric balloon operating at 30--40~km altitude. At these 
altitudes, more than 99\% of the atmospheric mass lies below the 
platform, substantially reducing absorption, scattering, and turbulence 
relative to ground-based systems, while the platform remains fully 
recoverable and reusable. Zero-pressure 
or super-pressure balloon envelopes may be used depending on the required 
flight duration: from multi-hour demonstration flights to multi-day 
campaigns in quasi-steady stratospheric wind patterns. The gondola houses 
the laser payload, telescope, pointing mechanisms, and power conditioning 
units, with energy supplied by onboard batteries dimensioned to the 
mission profile.

Stratospheric winds cause horizontal drift at typical speeds of 
10--30~m/s, which must be accounted for in mission planning but does 
not constitute a fundamental constraint: Earth's rotation provides 
access to all orbital inclinations below the launch latitude regardless 
of east-west drift, and launch site selection at low or mid-latitudes 
can optimise overflight geometry with respect to targeted orbital 
shells.

\subsection{Debris Engagement Sequence}

The engagement sequence follows the general architecture proposed 
for ground-based and space-based laser systems, adapted here to the specific constraints of a 
stratospheric balloon platform. An initial orbit determination 
from ground-based radar provides position uncertainties of several tens of metres, sufficient to cue the gondola when a target is expected to transit its field of regard. A mechanically actuated 
coarse-pointing stage slews toward the predicted line-of-sight, 
after which an onboard telescope and fast-steering mirror (FSM) 
system refines pointing and compensates residual gondola motion.

Two motion disturbances specific to the stratospheric platform
must be addressed. The horizontal drift at 10--30~m/s
introduces a platform velocity that, at a representative
engagement distance of 500~km, produces a target angular rate
of $\sim$20--60~$\mu$rad/s (0.02--0.06~mrad/s) --- more than two
orders of magnitude smaller than the debris apparent angular
motion ($v_\mathrm{circ}/R \approx 7.5~\mathrm{km/s}/500~\mathrm{km}
\approx 15~\mathrm{mrad/s}$) but non-negligible in the pointing budget.
Second, low-frequency gondola pendulation (typically 0.01--0.1~Hz, 
amplitude 1--5°) must be actively compensated by the FSM 
throughout the engagement. These two disturbances set the 
bandwidth and stroke requirements for the FSM, which are 
distinct from those of fixed ground-based or inertially 
stabilised space-based platforms.

At 500~km range, a 15~cm object subtends $\sim$0.3~$\mu$rad, 
consistent with the angular precision achievable with 
state-of-the-art FSM technology. Once acquired, continuous 
image-based tracking is maintained and the laser beam is 
delivered coaxially with the tracking axis to minimise parallax. 
The engagement geometry --- irradiating from below at oblique 
incidence --- is analyzed in more detail in Section 4.4.

\subsection{Payload Architectures}

Three payload configurations are considered (Figure~\ref{fig4}, bottom), 
reflecting different trade-offs between airborne complexity, ground 
infrastructure, and scalability:

\textbf{Architecture A --- Laser source fully on-board:} The complete 
laser system, including harmonic conversion to UV, is integrated on 
the gondola. This minimises operational complexity and eliminates 
atmospheric uplink losses, but places the most demanding constraints 
on gondola mass, volume, thermal management, and electrical power.

\textbf{Architecture B --- Ground laser source with on-board UV 
conversion:} An infrared laser on the ground transmits a collimated 
beam to the balloon, where a frequency-conversion module generates UV 
for delivery to the debris. This extends the relay mirror concept of 
Martin et al.\ \cite{martin2009} to an autonomous stratospheric 
platform, and relaxes the airborne power and laser mass constraints 
at the cost of uplink propagation losses and increased system 
complexity.

\textbf{Architecture C --- Distributed two-balloon configuration:} 
The laser source and beam director are hosted on separate balloon 
platforms operating in loose coordination. This allows the mass and 
power budget to be distributed across two payloads, each within 
current balloon capability, at the cost of inter-balloon pointing 
and synchronisation.

\subsection{Optical Performance and Architecture Sizing}

A key performance advantage of the stratospheric platform is the 
ability to operate at UV wavelengths, where ablation coupling 
efficiency is substantially higher than in the infrared 
\cite{phipps2014}. For a representative debris size of 
$\sim$15~cm, a focal spot of comparable dimension is required 
at the target. Assuming diffraction-limited propagation in vacuum:

\begin{equation}
w \simeq \frac{M^{2}\lambda L}{\pi w_{0}},
\label{eq:spot}
\end{equation}

where $L$ is the propagation distance, $\lambda$ the laser 
wavelength, $M^{2}$ the beam quality factor, and 
$w_{0} \simeq 0.5D$ the beam waist at transmitting aperture
diameter $D$. At $\lambda = 1064$~nm, achieving a 15~cm spot
at 800~km would require $D \gtrsim 11$~m --- impractical for any
balloon-borne system. At $\lambda = 355$~nm, apertures of 
1.4--2.5~m are sufficient, as shown in the next paragraphs, motivating UV operation as the 
baseline for all three architectures; Figure~\ref{fig6} illustrates the associated focusing trade-off.

\paragraph{Spot size, coupling fraction and required energy}
When the beam spot $w$ exceeds the debris radius $r_d = 7.5$~cm, 
only a fraction of the laser energy intercepts the target. 
For a Gaussian beam, the coupling fraction is:

\begin{equation}
f_\mathrm{coup} = 1 - \exp\!\left(-\frac{2r_d^2}{w^2}\right).
\label{eq:fcoup}
\end{equation}

The laser output energy required to deliver ablation-threshold 
fluence $\Phi_\mathrm{th}$ to the debris surface is:

\begin{equation}
E_\mathrm{laser} = \frac{\Phi_\mathrm{th} \cdot A_d}
{f_\mathrm{coup} \cdot \eta_\mathrm{opt}},
\label{eq:elaser}
\end{equation}

where $A_d = \pi r_d^2 = 1.77\times10^{-2}$~m$^2$ and 
$\eta_\mathrm{opt} \approx 0.85$ accounts for optical 
transmission losses. Primary mirrors up to 2~m have been 
demonstrated on stratospheric balloon platforms \cite{blast06}, 
and the BLAST-TNG mission has extended this to 2.5~m using 
lightweight mirror fabrication techniques \cite{blasttng}, 
within a payload compatible with current super-pressure balloon 
capabilities of 2--3~tonnes. Apertures beyond this size would 
place prohibitive demands on gondola structural mass and volume; 
the three aperture values in Table~\ref{tab:spot} (1.4, 2.0 
and 2.5~m) therefore span the practical envelope for 
stratospheric deployment. For comparison, at $\lambda = 1064$~nm 
the same apertures yield spot radii of 31--55~cm at zenith ---
coupling fractions below 0.11 --- rendering IR ablation
effectively infeasible for 15~cm debris regardless of pulse 
energy, and confirming UV operation as the only viable 
wavelength regime from a stratospheric platform.

\paragraph{Impulse components and optimal elevation angle}

The following analysis rests on several simplifying 
assumptions that are stated explicitly: (i)~the debris 
orbit is circular at altitude $h_d$; (ii)~the applied 
impulse is small ($\Delta v \ll v_\mathrm{circ}$), 
permitting first-order linearisation of the orbital 
elements; (iii)~the impulse is treated as instantaneous 
and applied at a single point, rather than distributed 
over the engagement window; (iv)~the debris passes 
directly overhead (no cross-track offset, so no 
out-of-plane component); (v)~the balloon is stationary 
during the engagement; and (vi)~the momentum coupling 
coefficient $C_m$, geometric efficiency $\eta_c$, and 
optical transmission $\eta_\mathrm{opt}$ are independent 
of elevation angle.
The ablation plume also has a finite effective exhaust velocity, so a large $C_M$ cannot be interpreted as an arbitrarily high specific impulse. The present study therefore treats $\Delta v$ as an accumulated, multi-shot quantity: if achieving the required increment would remove a non-negligible fraction of the target mass, fragmentation risk must be assessed with material-specific models and the case is treated as collision-avoidance nudging rather than guaranteed removal. No such marginal manoeuvre is credited as a removed object in the throughput model.

The laser impulse on the debris has two components 
relevant to orbit modification. The 
\emph{retrograde} component ($\propto\cos\theta$) 
acts against the orbital velocity, directly reducing 
the orbital energy. The \emph{radial outward} 
component ($\propto\sin\theta$) acts along the 
Earth--debris direction, conserving angular momentum 
while modifying the orbit shape.

\textit{Radial impulse.}
For a small radial outward impulse $\Delta v_r$ applied 
to a circular orbit of radius $r$, the angular momentum 
$h = rv_\mathrm{circ}$ is conserved (the radial 
direction is perpendicular to $\mathbf{r}\times\mathbf{v}$). 
The new specific orbital energy is:

\begin{equation}
\varepsilon' = \frac{v_\mathrm{circ}^2 + \Delta v_r^2}{2} 
- \frac{\mu}{r} \approx -\frac{\mu}{2r} + 
\frac{\Delta v_r^2}{2},
\end{equation}

giving a new semi-major axis $a' \approx r$ to first 
order. The new eccentricity follows from 
$e'^2 = 1 + 2\varepsilon'h'^2/\mu^2$:

\begin{equation}
e' = \sqrt{\frac{r\,\Delta v_r^2}{\mu}} = 
\frac{\Delta v_r}{v_\mathrm{circ}}.
\end{equation}

Since the application point has $v_r \neq 0$, it is 
not an apse of the new orbit. Using $r_p = a'(1-e')$:

\begin{equation}
\Delta r_p^\mathrm{rad} = r - r_p \approx 
\frac{r\,\Delta v_r}{v_\mathrm{circ}}.
\label{eq:drp_rad}
\end{equation}

\textit{Retrograde impulse.}
For a small retrograde impulse $\Delta v$ the new 
tangential velocity is $v_t = v_\mathrm{circ} - \Delta v$, 
with $v_r = 0$ at the application point. Since 
$v_t < v_\mathrm{circ}$ (the circular velocity at $r$), 
the debris is moving too slowly to maintain the circular 
orbit and the trajectory curves inward: the application 
point becomes the \emph{apogee} of the new ellipse, 
$r_a = r$. The new semi-major axis is:

\begin{equation}
a' = -\frac{\mu}{2\varepsilon'} \approx 
r\!\left(1 - \frac{2\Delta v}{v_\mathrm{circ}}\right),
\end{equation}

and from $r_a = a'(1+e')$:

\begin{equation}
e' = \frac{r}{a'} - 1 \approx \frac{2\Delta v}
{v_\mathrm{circ}}.
\end{equation}

The perigee is then $r_p = 2a' - r_a = 2a' - r$:

\begin{equation}
\Delta r_p^\mathrm{ret} = r - r_p = 
r - (2a' - r) = 2(r - a') \approx 
\frac{4r\,\Delta v}{v_\mathrm{circ}}.
\label{eq:drp_ret}
\end{equation}

Comparing Eqs.~(\ref{eq:drp_rad}) and~(\ref{eq:drp_ret}), 
the radial component contributes to perigee lowering 
with \textbf{one quarter} the efficiency of an 
equal retrograde impulse. The key distinction is 
that a retrograde impulse simultaneously lowers 
the semi-major axis \emph{and} raises the 
eccentricity, whereas a radial impulse only 
raises the eccentricity (leaving $a$ unchanged 
to first order).

\begin{table*}[ht]
\centering
\small
\setlength{\tabcolsep}{5pt}
\begin{tabular}{cclcccccc}
\hline
$\lambda$ & $D$ (m) & $R$ (km) & $\theta$ & $M^2$ & 
$w$ (cm) & $f_\mathrm{coup}$ & $E_\mathrm{laser}$ (kJ) \\
\hline
\multirow{8}{*}{355~nm}
 & \multirow{3}{*}{1.4} 
   & 765 & 90° & 1.5      & 18.5 & 0.28 & 0.74--2.2 \\
 & & 765 & 90° & 2.0$^*$  & 24.7 & 0.17 & 1.2--3.6  \\
 & & 830 & 67° & 1.5      & 20.1 & 0.25 & 0.83--2.5 \\
\cline{2-8}
 & \multirow{2}{*}{2.0} 
   & 765 & 90° & 1.5 & 13.0 & 0.46 & 0.45--1.4 \\
 & & 830 & 67° & 1.5 & 14.1 & 0.41 & 0.51--1.5 \\
\cline{2-8}
 & \multirow{2}{*}{2.5} 
   & 765 & 90° & 1.5 & 10.4 & 0.62 & 0.34--1.0 \\
 & & 830 & 67° & 1.5 & 11.3 & 0.57 & 0.36--1.1 \\
\hline
\multirow{3}{*}{1064~nm$^\dagger$}
 & 1.4 & 765 & 90° & 1.5 & 55.4 & 0.02 & 10--31 \\
 & 2.0 & 765 & 90° & 1.5 & 38.8 & 0.07 & 3.0--8.9 \\
 & 2.5 & 765 & 90° & 1.5 & 31.0 & 0.11 & 1.9--5.7 \\
\hline
\end{tabular}
\caption{Spot radius $w$, coupling fraction 
$f_\mathrm{coup} = 1-\exp(-2r_d^2/w^2)$ for debris radius 
$r_d = 7.5$~cm, and required laser output energy for 
$\Phi_\mathrm{th} = 1$--$3$~J/cm$^2$ \cite{phipps2014} and 
$\eta_\mathrm{opt} = 0.85$. All rows at $M^2 = 1.5$ except 
$^*$($M^2 = 2$, shown as reference for beam quality 
degradation). The three UV aperture values span the practical 
range for stratospheric deployment within current super-pressure 
balloon payload constraints \cite{blast06, blasttng}. 
$^\dagger$At 1064~nm, spot radii are 3$\times$ larger than at 
355~nm, resulting in coupling fractions below 0.11 and required 
energies 6--14$\times$ higher than the UV equivalent, 
confirming that IR operation is impractical for this application.}
\label{tab:spot}
\end{table*}

\paragraph{Engagement geometry and perigee-lowering efficiency}
For a balloon at $h_b = 35$~km and debris at $h_d = 800$~km, 
the slant range as a function of elevation angle $\theta$ is:

\begin{equation}
R(\theta) = \frac{h_d - h_b}{\sin\theta} = \frac{765~\mathrm{km}}
{\sin\theta}.
\label{eq:range}
\end{equation}

The minimum range of 765~km occurs at zenith ($\theta = 90^\circ$); 
at $\theta = 45^\circ$ the range increases to $\sim$1080~km, and at 
$\theta = 30^\circ$ to $\sim$1530~km.
At elevation angle $\theta$, a total velocity 
increment $\Delta v$ decomposes as $\Delta v\cos\theta$ 
retrograde and $\Delta v\sin\theta$ radial. The 
combined perigee lowering per unit impulse is:

\begin{equation}
\frac{\Delta r_p}{\Delta v} = \frac{r}{v_\mathrm{circ}}
\left(4\cos\theta + \sin\theta\right).
\label{eq:drp_comb}
\end{equation}

Since $\Delta v \propto f_\mathrm{coup}(\theta)\cdot 
E_\mathrm{laser}$ for fixed laser output energy, 
the perigee-lowering efficiency per unit laser 
energy is:

\begin{equation}
f_\mathrm{eff}(\theta) \propto 
\left(4\cos\theta + \sin\theta\right)\cdot 
f_\mathrm{coup}(\theta),
\label{eq:feff_exact}
\end{equation}

where the coupling fraction and beam spot radius are:

\begin{align}
f_\mathrm{coup}(\theta) &= 1 - \exp\!\left(
-\frac{2r_d^2}{w^2(\theta)}\right), \\
w(\theta) &= \frac{M^2\lambda\,R(\theta)}{\pi w_0}, 
\end{align}

captures the range dependence through the beam 
spot radius $w(\theta)$ at the target, for debris 
radius $r_d$ and transmitting aperture 
$w_0 = D/2$. Equation~(\ref{eq:feff_exact}) is 
the exact form under the assumptions stated above.

The optimal elevation angle $\theta_\mathrm{opt}$ that maximises
$f_\mathrm{eff}$ depends on the aperture $D$ through
$w_0$: larger apertures produce smaller spots at a
given range, increasing $f_\mathrm{coup}$ and shifting
the balance between the angular efficiency factor
$(4\cos\theta + \sin\theta)$ and the coupling term.
No closed-form solution exists for Eq.~(\ref{eq:feff_exact});
solved numerically, $f_\mathrm{eff}(\theta)$ peaks at
$\theta_\mathrm{opt} \approx 56$--$60^\circ$ over the aperture
range considered (Figure~\ref{fig:feff}), approaching the
$61.5^\circ$ of the $w \gg r_d$ limit derived below. This
impulse-optimal angle is distinct from the energy-sizing
elevation $\theta^* = 67^\circ$ used to dimension the system
(Table~\ref{tab:arch}), which is set by the maximum slant range
$R^* = 830$~km at which threshold fluence is delivered rather
than by maximising $f_\mathrm{eff}$.

In the limiting case $w \gg r_d$ (spot radius much 
larger than debris radius), the coupling fraction 
simplifies to $f_\mathrm{coup} \approx 2r_d^2/w^2 
\propto \sin^2\theta$, yielding the analytical 
approximation:

\begin{equation}
f_\mathrm{eff}(\theta) \propto 
\left(4\cos\theta + \sin\theta\right)\cdot\sin^2\theta,
\label{eq:feff_approx}
\end{equation}

which has the closed-form optimum
$\theta_\mathrm{opt} \approx 61.5^\circ$ ($R(\theta_\mathrm{opt}) \approx 950$~km).
However, as shown in Table~\ref{tab:spot}, the 
coupling fractions for the architectures considered 
here range from 0.25 to 0.57, placing them outside 
the regime of validity of this approximation 
($w/r_d \lesssim 3$ for all cases). 
Equation~(\ref{eq:feff_approx}) is therefore
retained only as an analytical reference; all
design points use the exact expression
(\ref{eq:feff_exact}).

\begin{figure*}[t]
\centering
\includegraphics[width=0.92\textwidth]{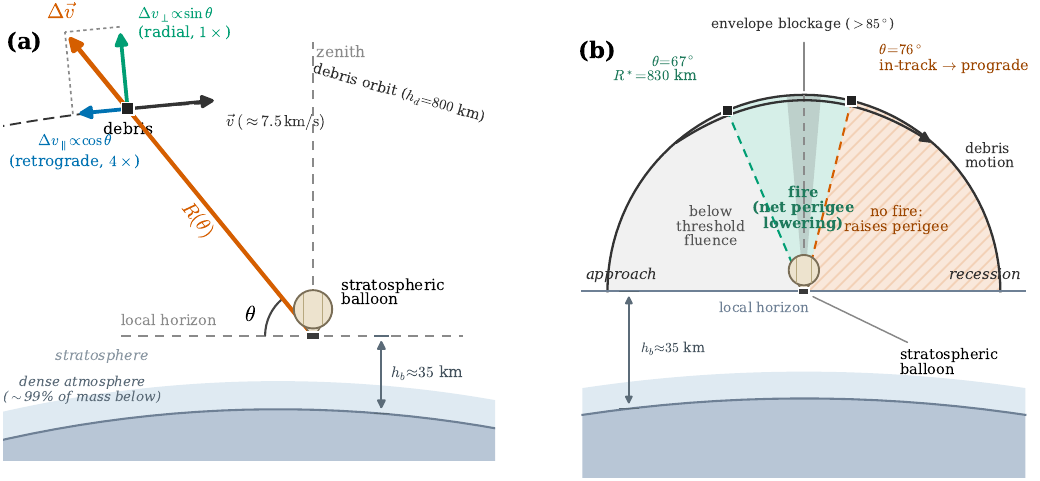}
\caption{\textbf{Engagement geometry and firing window.}
(a)~The ablation recoil is directed along the line of sight
(slant range $R(\theta)$) and decomposes into a retrograde in-track
component $\Delta v_\parallel \propto \cos\theta$ and a radial (outward)
component $\Delta v_\perp \propto \sin\theta$; per
Eqs.~(\ref{eq:drp_rad})--(\ref{eq:drp_ret}) the radial term lowers the
perigee with one quarter the efficiency of the retrograde term, giving
$\Delta r_p/\Delta v = (r/v_\mathrm{circ})(4\cos\theta + \sin\theta)$
(Eq.~\ref{eq:drp_comb}). (b)~On approach the in-track component is
retrograde; past zenith it turns prograde, and net perigee lowering
persists only while $\tan\theta > 4$ ($\theta \gtrsim 76^\circ$). The
usable window is therefore asymmetric --- $67^\circ$ on approach through
zenith to $\sim 76^\circ$ on recession --- and is further reduced by
balloon-envelope blockage above $\sim 85^\circ$. The platform operates
at $h_b \approx 35$~km, well above the dense lower atmosphere.}
\label{fig:geometry}
\end{figure*}

\begin{figure}[t]
\centering
\includegraphics[width=\columnwidth]{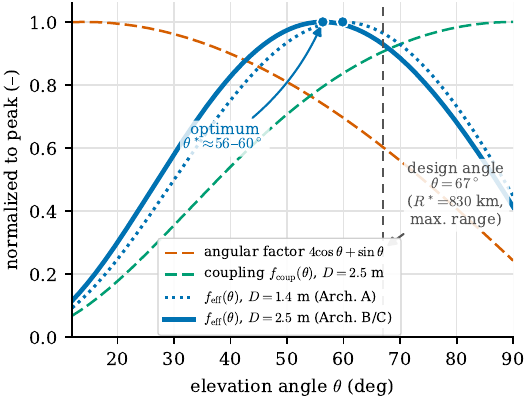}
\caption{\textbf{Perigee-lowering efficiency $f_\mathrm{eff}(\theta)$.}
$f_\mathrm{eff}(\theta) = (4\cos\theta + \sin\theta)\,f_\mathrm{coup}(\theta)$
(each curve normalised to its own peak). The decreasing angular factor and
the increasing coupling fraction yield an interior optimum at
$\theta_\mathrm{opt} \approx 56$--$60^\circ$ (aperture-dependent;
$61.5^\circ$ in the $w \gg r_d$ limit, Eq.~\ref{eq:feff_approx}). This
impulse optimum is distinct from the $67^\circ$ energy-sizing angle, which
is set by the maximum slant range $R^* = 830$~km at which threshold fluence
is delivered.}
\label{fig:feff}
\end{figure}

In practice, the irradiation interval is
asymmetric about zenith. On the approach
branch the in-track component is retrograde
and lowers the perigee; past zenith it becomes
prograde, and the net perigee change (radial
minus in-track) remains negative only while
$\tan\theta > 4$, i.e.\ for $\theta \gtrsim 76^\circ$
on recession. Below that elevation on the receding
branch, continued irradiation would raise rather than
lower the perigee. The useful arc therefore runs from
$\theta = 67^\circ$ on approach, through zenith, down
to $\sim 76^\circ$ on recession (further curtailed by
envelope blockage above $\sim 85^\circ$), as illustrated
in Figure~\ref{fig:geometry}(b).

\subsubsection*{\textbf{Architecture A --- UV laser fully on-board}}

The complete DPSSL system including THG, beam director, 
battery, thermal management, and all support systems are 
integrated on a single gondola within the $\sim$2{,}500~kg 
payload capacity of current super-pressure balloons 
\cite{blast06}. The aperture selection is driven by both 
the mass budget and the gondola mechanical design.

From a mass perspective, the laser system --- comprising 
the gain medium, pump diode arrays, power electronics, and 
THG module --- is the dominant mass driver and the primary 
technology challenge of this architecture. A kJ-class UV 
pulsed laser system at 355~nm does not yet exist in a 
compact form; current DPSSL systems in this energy class 
operate at infrared wavelengths and occupy laboratory-scale 
installations (TRL~2--3 for the UV kJ-class configuration, 
Table~\ref{tab:trl}). A mass estimate of 1~ton is 
therefore projected based on scaling from existing DPSSL 
architectures and assumes 
continued progress in diode pumping efficiency, nonlinear 
conversion, and thermal management --- precisely the 
technology development that the STRATOLASER project aims 
to characterise \cite{stratolaser2025}. This uncertainty 
is the principal risk of Architecture~A and motivates the 
alternative configurations described later.

Thermal management of the laser is the second major mass 
driver. During the 70~s engagement window, waste heat is 
generated at:

\begin{equation}
\dot{Q}_\mathrm{waste} = P_\mathrm{elec} \cdot 
(1 - \eta_{e\text{-}o}) = 88 \times 0.83 
\approx 73~\mathrm{kW},
\end{equation}

accumulated over an engagement as $Q_\mathrm{eng} = 
73 \times 70 \approx 5{,}100$~kJ. At stratospheric 
altitude convective cooling is negligible ($\rho \approx 
0.5\%$ of sea level), so rejection is purely radiative. 
The net radiative power at an operating temperature of
350~K against a stratospheric ambient of $\sim$218~K is
$\sigma\varepsilon(T^4 - T_\mathrm{amb}^4) \approx
650$~W/m$^2$ (assuming a radiator emissivity
$\varepsilon \approx 0.9$). Allocating $\sim$400~kg to deployable 
radiator panels at 10~kg/m$^2$ yields an area of 
$\sim$40~m$^2$ and a continuous rejection capacity of 
$\sim$26~kW. The net heat stored per engagement 
(accounting for simultaneous rejection during firing) is 
$Q_\mathrm{stored} \approx (73-26)\times70 \approx 
3{,}300$~kJ, requiring a post-engagement cooling interval 
of $t_\mathrm{cool} \approx 3{,}300/26 \approx 127$~s 
before the next engagement. This gives an operational 
cadence of $\sim$70 + 127 $\approx$ 200~s per debris 
object, permitting up to $\sim$100 engagements over a 
6-hour flight.

The battery and power conditioning system must satisfy two 
simultaneous constraints. The energy constraint arises from 
supplying up to 100 engagement bursts of 70~s during a 
6-hour flight at $P_\mathrm{elec} = 88$~kW, plus 3~kW 
continuous auxiliary power, totalling:

\begin{equation}
E_\mathrm{total} = \frac{88 \times 70 \times 100}{3{,}600} 
+ 3 \times 6 \approx 171 + 18 = 189~\mathrm{kWh}.
\end{equation}

At 250~Wh/kg this corresponds to $\sim$756~kg for the 
battery cells alone, which is prohibitive. In practice, 
the number of engagements is limited by the debris 
population accessible during a single flight and by the 
thermal cycling constraints of the laser; a more 
representative operational scenario of 15 engagements 
per flight gives:

\begin{equation}
E_\mathrm{total} \approx \frac{88 \times 70 \times 15}
{3{,}600} + 18 \approx 25 + 18 = 43~\mathrm{kWh},
\end{equation}

corresponding to $\sim$172~kg at 250~Wh/kg. The peak 
power constraint of 88~kW is satisfied by high-rate 
lithium cells at $\sim$2~kW/kg (44~kg), so the energy 
constraint dominates. Including power conditioning 
electronics, current drivers for the pump diode arrays, 
and battery thermal management, the battery subsystem 
is estimated at $\sim$250~kg.

Combining all subsystems --- laser, radiators, battery, 
telescope and pan-and-tilt assembly, gondola structure, 
electronics, and recovery hardware --- the estimated total
gondola mass is approximately 2{,}300--2{,}600~kg, at or
above the $\sim$2{,}500~kg payload limit at the upper end. The detailed breakdown is 
provided in the comparative Table~\ref{tab:arch}.

From a mechanical design perspective, the beam director 
is mounted on a pan-and-tilt system at the base of the 
gondola pointing upward, while the laser source and heavy 
subsystems are fixed above, attached to the main balloon 
suspension mast. This configuration constrains the 
maximum practical mirror aperture: a larger mirror sweeps 
a greater volume during slewing, increasing the moment of 
inertia and the amplitude of pendulation-induced 
disturbances, which in turn require increasingly massive 
actuators, counterweights, and active compensation 
systems. Increasing the primary mirror from $D = 1.4$~m 
to $D = 2.0$~m would add approximately 250--350~kg 
through the combined effect of mirror mass, telescope 
structure, and the scaled pan-and-tilt compensation 
hardware, pushing the total gondola mass beyond 
2{,}500~kg. The $D = 1.4$~m aperture is therefore 
selected as the baseline for Architecture~A; in 
Architectures~B and~C the laser is removed from the 
gondola, freeing 500--800~kg and permitting larger 
apertures.

The system is dimensioned at the design elevation angle 
$\theta^* = 67^\circ$, corresponding to the maximum slant range 
$R^* = 830$~km at which ablation threshold fluence is 
delivered. Since the slant range is symmetric about zenith, 
the engagement window extends from $\theta = 67^\circ$ on 
approach through zenith to $\theta = 67^\circ$ on recession, 
spanning a total of 46° in elevation. The apparent angular 
velocity of the debris as seen from the balloon is:

\begin{equation}
\dot{\theta} \approx \frac{v_\mathrm{circ}}{R^*} = 
\frac{7{,}460~\mathrm{m/s}}{830\times10^3~\mathrm{m}} 
\approx 0.51^\circ/\mathrm{s},
\end{equation}

giving a total firing window of:

\begin{equation}
t_\mathrm{window} = \frac{46^\circ}{0.51^\circ/\mathrm{s}} \approx 
90~\mathrm{s}.
\end{equation}

However, a further geometric constraint specific to the balloon 
platform is that the balloon envelope itself occludes 
the field of view near zenith, blocking elevations 
approximately above $\theta \approx 85^\circ$. This reduces
the usable firing window from the theoretical
$2\times23^\circ = 46^\circ$ to $2\times18^\circ = 36^\circ$, providing a total firing window of 70~s.
This $\sim$70~s value is the geometric, fluence-limited engagement
time (symmetric about zenith) that sets the thermal duty cycle;
as shown in Figure~\ref{fig:geometry}(b), only the pre-zenith branch
and the near-zenith recession arc down to $\theta \approx 76^\circ$
contribute net perigee lowering, so a per-pass $\Delta v$ accumulation
would use a smaller effective arc than the full geometric window.

Alternatively, targeting passes whose peak elevation 
is $\sim$85° rather than 90° avoids the blockage 
entirely. However, such passes introduce a cross-track 
velocity component in addition to the retrograde and 
radial components, as the debris trajectory is no longer 
contained in the vertical plane through the balloon. 
The fraction of the impulse directed cross-track 
represents a loss that does not contribute to perigee 
lowering; at a peak elevation offset of $5^\circ$ from 
zenith this loss is $\sim\sin(5^\circ) \approx 8.7\%$ 
at peak, and less when averaged over the pass. 
Both operational modes are viable;   The transmittance curves in Figure~\ref{fig5} are for a zenith path (airmass $X=1$). For the usable elevation range $67^\circ\le\theta\le85^\circ$, a plane-parallel approximation gives $X\simeq1/\sin\theta$, i.e., 1.09 at the worst-case $67^\circ$ and 1.00 at $85^\circ$. We therefore apply the minimum transmittance at $\theta=67^\circ$ as the conservative sizing case; curved-atmosphere and weather variability are left to detailed   design.
The overall wall-plug efficiency is $\eta_{e\text{-}o} 
= \eta_\mathrm{diode} \cdot \eta_\mathrm{THG} \approx 
0.52 \times 0.33 \approx 0.17$. For $E_\mathrm{laser} 
= 0.83$--$2.5$~kJ per pulse at 10~Hz (from 
Table~\ref{tab:spot}, $D = 1.4$~m, $R = 830$~km, 
$M^2 = 1.5$), the instantaneous electrical power 
during a burst is $P_\mathrm{elec} = 49$--$147$~kW.

The energy deposited on the debris per pulse is 
$E_\mathrm{dep} = f_\mathrm{coup} \cdot \eta_\mathrm{opt} 
\cdot E_\mathrm{laser} = 176$--$531$~J, yielding a 
velocity increment:

\begin{equation}
\Delta v = \frac{C_m \cdot E_\mathrm{dep} \cdot \eta_c}
{\mu \cdot A_d} \approx 0.5\text{--}1.5~\mathrm{m/s~per~pulse},
\end{equation}

where $C_m \approx 100$~N/MW at 355~nm \cite{phipps2014}, 
$\eta_c \approx 0.5$, $\mu = 1$~kg/m$^2$, and 
$A_d = 1.77\times10^{-2}$~m$^2$.
The combined retrograde and radial efficiency factor at
$\theta^* = 67^\circ$, consistent with
Eq.~(\ref{eq:drp_comb}), is
$f_\mathrm{geo} = \cos\theta^* + \tfrac{1}{4}\sin\theta^*
\approx 0.62$ (the radial term enters at one-quarter
weight; Eqs.~(\ref{eq:drp_rad})--(\ref{eq:drp_ret})),
giving an effective perigee-lowering $\Delta v$ of
$\approx 0.3$--$0.9$~m/s per pulse. A full assessment 
of multi-pass mission performance requires orbit 
propagation and is outside the scope of this 
conceptual study; the $\Delta v$ per pulse and the 
$\sim$70~s engagement window per pass are the primary 
figures of merit used for architecture comparison.

The baseline target is a 15~cm diameter, thin metallic fragment represented by an areal density $\mu=1$~kg/m$^2$ (with 1--5~kg/m$^2$ sensitivity in the cost model); this is a conservative geometric proxy rather than a claim about a unique debris shape or material.

\subsubsection*{\textbf{Architecture B --- Ground IR laser with 
independent on-board UV conversion}}

Removing the laser source from the gondola eliminates the 
three dominant mass drivers of Architecture~A: the DPSSL 
laser head, pump diodes and power electronics 
($\sim$550--850~kg), the associated 400~kg radiator array, 
and the 250~kg high-power battery. What remains on-board 
from the laser chain is only the THG conversion module 
($\sim$50~kg) and a dedicated IR receiving aperture 
($\sim$30~kg). This saving of $\sim$1{,}200~kg relative 
to Architecture~A --- combined with the substantially 
reduced battery requirement, since the balloon need only 
power pointing, electronics, and THG control 
($\sim$3.5~kW) --- allows the primary transmitting 
mirror to be increased to $D = 2.5$~m, matching the 
aperture of Architecture~C, while the total gondola 
mass remains at $\sim$1{,}760~kg, well within the 
payload limit.

A ground-based Yb:YAG DPSSL ($\lambda = 1030$~nm) 
transmits a collimated beam through a 0.5~m aperture 
to the balloon at 35~km altitude. The Rayleigh range 
of the beam is $z_R = \pi w_0^2/\lambda \approx 190$~km, 
far exceeding the 35~km uplink distance, so diffraction 
spreading is negligible and the beam arrives at the 
balloon with essentially the same 0.5~m diameter. 
Since the beam aperture ($D = 0.5$~m) is comparable 
to the atmospheric coherence length at 1030~nm 
($r_0 \approx 0.54$~m under typical conditions), 
the uplink wavefront is nearly diffraction-limited 
without adaptive optics correction, and AO is not 
a strict requirement. A dedicated 0.6~m receiving 
aperture, mounted independently from the 2.5~m 
transmitting telescope, collects the uplink beam. 
The uplink efficiency budget is:

\begin{equation}
\eta_\mathrm{uplink} = \eta_\mathrm{atm} \cdot 
\eta_\mathrm{AO} \cdot \eta_\mathrm{coupling} 
\approx 0.87 \times 0.75 \times 0.90 \approx 0.59,
\end{equation}

where $\eta_\mathrm{atm} \approx 0.87$ at 1030~nm 
for clear conditions, $\eta_\mathrm{AO} \approx 0.75$ 
represents residual wavefront losses (conservative 
estimate; AO correction is optional at this aperture), 
and $\eta_\mathrm{coupling} \approx 0.90$ is the 
geometric coupling into the receiver. Including THG 
conversion ($\eta_\mathrm{THG} \approx 0.32$), 
the overall ground-to-UV efficiency is:

\begin{equation}
\eta_\mathrm{ground\to UV} = \eta_\mathrm{uplink} 
\cdot \eta_\mathrm{THG} \approx 0.59 \times 0.32 
\approx 0.19.
\end{equation}

To deliver the UV pulse energy required by the 2.5~m 
transmitting telescope ($E_\mathrm{laser} = 0.7$~kJ 
nominal, Table~\ref{tab:spot}, $D = 2.5$~m, 
$R = 830$~km, $M^2 = 1.5$), the required ground 
laser output is:

\begin{equation}
E_\mathrm{ground} = \frac{E_\mathrm{laser}}
{\eta_\mathrm{ground\to UV}} = 
\frac{0.7~\mathrm{kJ}}{0.19} \approx 3.7~\mathrm{kJ}.
\end{equation}

A ground-based 3--4~kJ Yb:YAG DPSSL at 10~Hz is 
within the scaling roadmap of the DiPOLE/BIVOJ 
architecture (TRL~4--5, Table~\ref{tab:trl}) 
\cite{mason2011, lorbeer2018}, benefits from 
full laboratory infrastructure for cooling and 
power, and represents a substantially more mature 
technology than the compact UV kJ-class laser 
required by Architecture~A.

The THG module dissipates $(1-\eta_\mathrm{THG}) 
= 68\%$ of the received IR power as waste heat. 
For the nominal ground output of 3.7~kJ at 10~Hz, 
the IR power received at the balloon is 
$3.7\times10\times0.59 \approx 21.8$~kW, producing 
a THG waste heat of:

\begin{equation}
\dot{Q}_\mathrm{THG} = 21.8 \times 0.68 \approx 
15~\mathrm{kW}.
\end{equation}

A radiator of 23~m$^2$ (230~kg at 10~kg/m$^2$) 
provides a continuous rejection capacity of 
$\sim$15~kW, matching the THG waste heat rate 
and enabling essentially continuous operation 
with negligible inter-engagement cooling. 
Unlike Architecture~A, the operational cadence 
is therefore not thermally limited but governed 
by atmospheric uplink availability: under clear 
sky conditions the cadence approaches the 
engagement window ($\sim$70~s), while cloud 
cover or strong turbulence can interrupt the 
IR uplink entirely. This weather dependence 
is the primary operational risk of 
Architecture~B and the key distinction 
from Architecture~C, which is fully 
weather-independent.

The gondola battery supplies only 
$\sim$3.5~kW for pointing, electronics, 
and THG control, requiring $\sim$21~kWh 
over a 6-hour flight and corresponding 
to $\sim$120~kg including power 
conditioning. The total gondola mass is 
estimated at $\sim$1{,}760~kg. The energy 
deposited on the debris and the resulting 
$\Delta v$ are identical to Architecture~C 
(same aperture and design range): 
$E_\mathrm{dep} \approx 339$~J and
$\Delta v \approx 0.96$~m/s per pulse,
with an effective perigee-lowering
increment of $f_\mathrm{geo}\Delta v \approx 0.60$~m/s
(with $f_\mathrm{geo}=0.62$).
The detailed mass breakdown is provided 
in Table~\ref{tab:arch}.

\subsubsection*{\textbf{Architecture C --- Distributed two-balloon 
configuration}}

Architecture~C separates the laser source and the beam director 
onto two independent balloon platforms operating at 35~km 
altitude in loose coordination. Gondola~1 hosts the UV laser 
source and its associated power and thermal subsystems; 
Gondola~2 hosts the 2.5~m beam director telescope, pointing 
system, and tracking electronics. This separation eliminates 
the fundamental mass and integration conflict of 
Architecture~A, where the laser competes with the precision 
optics for payload volume, structural allocation, and thermal 
isolation, and where the pan-and-tilt system must 
simultaneously manage the inertia of the laser and the 
telescope. In Gondola~2, the entire mechanical design is 
dedicated to the optical subsystem without compromise, 
enabling a 2.5~m primary mirror --- the largest aperture 
demonstrated on stratospheric balloon platforms 
\cite{blasttng} --- to be integrated with a pan-and-tilt 
and compensation system sized exclusively to its optical 
and inertial requirements.

The inter-balloon optical path at 35~km altitude is 
effectively in vacuum. For a typical separation of 10~km, 
the Rayleigh range of a 0.3~m beam at 355~nm is 
$z_R \approx 200$~km, far exceeding the inter-balloon 
distance, so diffraction losses are negligible. 
Differential drift  the two platforms is typically 
well below 0.1~m/s at the same altitude and latitude, 
corresponding to an angular rate of $\sim$0.01~mrad/s 
at 10~km separation --- well within FSM tracking 
capability. The inter-balloon synchronisation and beam 
transfer will be demonstrated during the STRATOLASER 
balloon-to-balloon experimental campaign planned for 
2027 \cite{stratolaser2025}.

\textit{Gondola 2 --- beam director.} From 
Table~\ref{tab:spot} ($D = 2.5$~m, $R = 830$~km, 
$M^2 = 1.5$), the required UV pulse energy at the 
telescope input is $E_\mathrm{laser} = 0.36$--$1.1$~kJ. 
Gondola~2 houses the telescope, pan-and-tilt assembly, 
FSMs, tracking electronics, UV injection optics, and 
an electronics-only battery ($\sim$6~kW for 6 hours: 
$\sim$150~kg), giving an estimated total mass of 
$\sim$1{,}300~kg.

\textit{Gondola 1 --- laser source.} Using 
$E_\mathrm{laser} = 0.7$~kJ as the nominal design point 
and $\eta_\mathrm{inter} \approx 0.95$, the required 
laser output is $E_\mathrm{G1} \approx 0.74$~kJ, 
with $\eta_{e\text{-}o} \approx 0.17$ giving:

\begin{equation}
P_\mathrm{elec} = \frac{0.74 \times 10}{0.17} \approx 
44~\mathrm{kW}, \quad 
\dot{Q}_\mathrm{waste} = 44 \times 0.83 \approx 36~\mathrm{kW}.
\end{equation}

Since Gondola~1 carries no telescope or precision optics, 
its full payload budget can be allocated to maximising 
radiator area and battery capacity. Allocating 500~kg 
to radiator panels (50~m$^2$ at 10~kg/m$^2$) yields a 
continuous rejection capacity of $\sim$32.5~kW, 
nearly matching the 36~kW waste heat rate. The net 
heat stored per 70~s engagement is therefore only:

\begin{equation}
Q_\mathrm{stored} = (36-32.5)\times 70 \approx 245~\mathrm{kJ},
\end{equation}

requiring a post-engagement cooling interval of just 
$t_\mathrm{cool} \approx 245/32.5 \approx 7.5$~s, 
giving an operational cadence of $70 + 7.5 \approx 
78$~s per debris object. Allocating the remaining 
payload budget ($\sim$1{,}000~kg) to battery and power 
conditioning provides approximately:

\begin{equation}
E_\mathrm{battery} \approx 1{,}000 \times 0.25 - 18 
\approx 232~\mathrm{kWh},
\end{equation}

sufficient for $\sim$270 engagements at 0.856~kWh per 
burst, consistent with the 78~s cadence over a 6-hour 
flight. Gondola~1's total estimated mass is 
$\sim$2{,}500~kg, fully utilising the payload limit 
--- in contrast to Architectures~A and~B where the 
telescope integration constrains the mass available 
for power and thermal subsystems.

The energy deposited on the debris per pulse is 
$E_\mathrm{dep} = 0.57 \times 0.85 \times 0.7 \approx 
339$~J, yielding:

\begin{align}
\Delta v &= \frac{C_m \cdot E_\mathrm{dep} \cdot \eta_c}
{\mu \cdot A_d} \nonumber \\
&= \frac{100\times10^{-6} \times 339 \times 0.5}
{1 \times 1.77\times10^{-2}} \approx 0.96~\mathrm{m/s~per~pulse,}
\end{align}

with an effective perigee-lowering increment of
$f_\mathrm{geo} \cdot \Delta v \approx 0.60$~m/s,
comparable to Architecture~A at less than half the
laser pulse energy. The detailed mass breakdown of 
both gondolas is provided in Table~\ref{tab:arch}.

Table~\ref{tab:trl} summarises the estimated TRL of the key 
subsystems. Table~\ref{tab:arch} provides a high-level numerical 
comparison of the three architectures.

\begin{table*}[ht]
\centering
\small
\setlength{\tabcolsep}{5pt}
\begin{tabular}{llcc}
\hline
\textbf{Subsystem / Technology} & \textbf{Arch.} & \textbf{TRL} & \textbf{Ref.} \\
\hline
\multicolumn{4}{l}{\textit{Platform (all architectures)}} \\
Super-pressure balloon, $>$100-day endurance, 2{,}500~kg payload & A,B,C & 8--9 & \cite{cathey2008} \\
Gondola pan-and-tilt for primary mirrors up to 2.0~m & A,B & 7--8 & \cite{blast06} \\
Gondola pan-and-tilt for primary mirrors up to 2.5~m & C & 5--6 & \cite{blasttng} \\
Active gondola pendulation damping, $<$1~arcmin residual & A,B,C & 6--7 & --- \\
\hline
\multicolumn{4}{l}{\textit{Sensing and tracking (all architectures)}} \\
Ground SSA radar, $<$100~m position uncertainty & A,B,C & 9 & --- \\
Non-cooperative RSO optical acquisition from stratosphere & A,B,C & 4--5 & \cite{chianelli2024} \\
Fast-steering mirror, $>$100~Hz bandwidth, $>$10~mrad stroke & A,B,C & 7--8 & --- \\
\hline
\multicolumn{4}{l}{\textit{Optics and beam control}} \\
UV-grade composite primary mirror, 1.4~m$^*$ & A & 6--7 & \cite{blast06} \\
UV-grade composite primary mirror, 2.0~m$^*$ & B & 5--6 & \cite{blast06} \\
UV-grade composite primary mirror, 2.5~m$^*$ & C & 4--5 & \cite{blasttng} \\
Compact airborne THG module, gondola-rated, $>$30\% eff., kJ-class$^{**}$ & A,B,C & 4--5 & \cite{aladin2009} \\
Coaxial beam director with non-cooperative debris tracking & A,B,C & 7--8 & \cite{lamberson2004} \\
Radiator panels, 12--50~m$^2$ & A,C & 9 & --- \\
\hline
\multicolumn{4}{l}{\textit{Laser source --- Architecture A}} \\
Compact UV DPSSL, 1.5~kJ at 355~nm, $<$1~ton total & A & 2--3 & \cite{stratolaser2025} \\
\hline
\multicolumn{4}{l}{\textit{Laser source and uplink --- Architecture B}} \\
Ground-based IR DPSSL, 5--8~kJ/pulse at 1064~nm, stationary & B & 6--7 & \cite{mason2011} \\
AO wavefront correction for 35~km uplink & B & 8--9 & \cite{eos} \\
\hline
\multicolumn{4}{l}{\textit{Laser source and inter-balloon link --- Architecture C}} \\
Compact gondola-rated UV DPSSL, 0.74~kJ/pulse at 355~nm (G1) & C & 2--3 & \cite{stratolaser2025} \\
Inter-balloon UV beam pointing and transfer, 1-10~km separation & C & 2--3 & \cite{stratolaser2025} \\
Dual-gondola flight synchronisation and telemetry & C & 9 & --- \\
\hline
\multicolumn{4}{l}{\textit{System level}} \\
Integrated stratospheric laser ablation system & A,B,C & 2 & \cite{stratolaser2025}\\
\hline
\end{tabular}
\caption{Estimated TRL of key STRATOLASER subsystems per ESA 
definitions. The STRATOLASER project targets TRL~4 for the 
integrated system \cite{stratolaser2025}.
$^*$UV-grade mirrors ($\lambda/10$ surface quality at 355~nm) 
differ substantially from the submillimetre-optimised mirrors 
flown on BLAST; demonstrated balloon mirrors achieve $\sim\mu$m 
roughness adequate for submillimetre but not UV operation, 
reducing the effective TRL relative to the balloon platform heritage.
$^{**}$THG modules at mJ-class pulse energy are space-proven 
(ESA Aeolus/ALADIN); the kJ-class airborne variant required 
here operates three orders of magnitude higher in pulse energy, 
where optical damage thresholds in nonlinear crystals become 
the limiting factor, reducing the TRL accordingly.
G2: Gondola~2 (beam director, Architecture~C).}
\label{tab:trl}
\end{table*}

\begin{table*}[!t]
\centering
\small
\setlength{\tabcolsep}{4pt}
\begin{tabular}{p{4.8cm}p{3.6cm}p{3.6cm}p{3.6cm}}
\hline
\textbf{Parameter} & 
\textbf{A: On-board UV} & 
\textbf{B: Ground IR + UV} & 
\textbf{C: Two-balloon} \\
\hline
\multicolumn{4}{l}{\textit{Optical}} \\
Transmitting aperture & 1.4~m & 2.5~m & 2.5~m (G2) \\
Receiving aperture & --- & 0.6~m (G2, IR uplink) & 0.3~m (G2, from G1)$^a$ \\
Wavelength: ground / on-board & --- / 355~nm & 1030~nm / 355~nm & --- / 355~nm \\
Design range / orbit$^b$ & 830~km / 800~km alt. & 830~km / 800~km alt. & 830~km / 800~km alt. \\
$f_\mathrm{coup}$$^c$ ($M^2=1.5$) & 0.25 & 0.57 & 0.57 \\
\hline
\multicolumn{4}{l}{\textit{Laser source}} \\
Laser energy: on-board / ground$^d$ & 1.5~kJ / --- & --- / 3.7~kJ & 0.74~kJ (G1) / --- \\
Transmission efficiency$^e$ & 1.0 (vacuum) & 0.19 (ground$\to$UV) & 0.95 (vacuum, G1$\to$G2) \\
$P_\mathrm{elec}$ during engagement & 88~kW & $<$5~kW (THG only) & 44~kW (G1) \\
On-board / ground laser TRL & 2--3 / --- & --- / 4--5 & 2--3 / --- \\
\hline
\multicolumn{4}{l}{\textit{Engagement performance}} \\
$E_\mathrm{dep}$ per pulse & 319~J & 339~J & 339~J \\
$\Delta v$ / effective $\Delta v$$^f$ & 0.90 / 0.56~m/s & 0.96 / 0.60~m/s & 0.96 / 0.60~m/s \\
Engagement window$^b$ & 70~s & 70~s & 70~s \\
\hline
\multicolumn{4}{l}{\textit{Thermal management}} \\
On-board waste heat & 73~kW & 15~kW (THG) & 36~kW (G1) \\
Radiator area / mass & 40~m$^2$ / 400~kg & 23~m$^2$ / 230~kg & 50~m$^2$ / 500~kg (G1) \\
Continuous rejection & 26~kW & 15~kW & 32.5~kW (G1) \\
Heat stored / cooling time & 3{,}290~kJ / 127~s & $\approx$0 / $\approx$0~s & 245~kJ / 7.5~s \\
\hline
\multicolumn{4}{l}{\textit{Operational performance}} \\
Cadence / engagements per 6h & $\sim$200~s / $\sim$110 & weather-dep.$^g$ / $\sim$270 max & $\sim$78~s / $\sim$270 \\
\hline
\multicolumn{4}{l}{\textit{Mass budget}} \\
Laser + THG + laser thermal (kg) & $\sim$700--1{,}000$^h$ & $\sim$80 (THG + receiver) & $\sim$500 (G1) \\
Radiators (kg) & 400 & 230 & 500 (G1) \\
Battery + power cond. (kg) & $\sim$250 & $\sim$120$^i$ & $\sim$1{,}000 (G1) \\
Telescope + pointing + FSM (kg) & $\sim$450 & $\sim$800 & $\sim$750 (G2) \\
Structure + recovery + misc (kg) & $\sim$500 & $\sim$530 & $\sim$350 each \\
\textbf{Total gondola mass} & $\sim$\textbf{2{,}300--2{,}600~kg} &
$\sim$\textbf{1{,}760~kg} &
\textbf{$\sim$2{,}500~kg (G1)} \\
& & & \textbf{$\sim$1{,}300~kg (G2)} \\
\hline
\multicolumn{4}{l}{\textit{Infrastructure and risk}} \\
Ground infrastructure & Radar & Radar + 3.7~kJ DPSSL & Radar \\
Key TRL bottleneck & Compact kJ UV (TRL~2--3) & kJ IR DPSSL at 10~Hz (TRL~4--5) & UV laser + 2.5~m pan-and-tilt \\
Key operational risk & On-board integration complexity & Weather disrupting IR uplink & 2.5~m pan-and-tilt mechanical operation \\
\hline
\end{tabular}
\caption{Comparative summary of the three payload architectures.
All mass and performance estimates are indicative.
$^a$Architecture~C: dedicated 0.3~m injection port on G2
receives inter-balloon UV beam from G1; 2.5~m mirror
transmits toward debris only.
$^b$Design range 830~km corresponds to $\theta^*=67^\circ$ for
an 800~km reference orbit (energy-sizing case; distinct from the
$f_\mathrm{eff}$ optimum $\theta_\mathrm{opt}\approx56$--$60^\circ$).
The geometric, fluence-limited firing window spans
$\theta=67^\circ$ pre-zenith to $\theta=67^\circ$
post-zenith ($\sim$90~s total); usable window $\sim$70~s
after excluding balloon zenith blockage ($\theta>85^\circ$).
$^c$$f_\mathrm{coup}=1-\exp(-2r_d^2/w^2)$: fraction of
energy intercepted by debris of radius $r_d=7.5$~cm.
$^d$Architecture~B ground laser energy accounts for
$\eta_\mathrm{total}=0.19$ (uplink + THG losses).
Architecture~C value is at G1 output
($\eta_\mathrm{inter}=0.95$).
$^e$Architecture~B: $\eta=\eta_\mathrm{atm}\cdot
\eta_\mathrm{AO}\cdot\eta_\mathrm{coupling}\cdot
\eta_\mathrm{THG}\approx0.19$.
$^f$$\Delta v$: $C_m=100$~N/MW, $\eta_c=0.5$,
$\mu=1$~kg/m$^2$ \cite{phipps2014};
$f_\mathrm{geo}=\cos67^\circ+\tfrac{1}{4}\sin67^\circ\approx0.62$
(radial term at one-quarter weight, consistent with
Eqs.~(\ref{eq:drp_rad})--(\ref{eq:drp_ret})).
$^g$Architecture~B cadence is not thermally limited
(laser on ground); operational availability is
governed by atmospheric uplink conditions
(cloud cover, turbulence) and may vary significantly.
Maximum cadence under clear-sky conditions
is comparable to Architecture~C.
$^h$Architecture~A laser mass is the dominant
uncertainty (500--1{,}000~kg); upper bound drives
gondola to the 2{,}500~kg payload limit, motivating
the 1.4~m aperture.
$^i$Architecture~B balloon battery powers pointing,
electronics and THG control only ($\sim$3.5~kW);
the laser power supply remains on the ground.
G1:~Gondola~1 (laser, Arch.~C);
G2:~Gondola~2 (beam director, Arch.~B~and~C).}
\label{tab:arch}
\end{table*}

\section{Techno-Economic Comparison}

\subsection{Architecture Comparison}

Table~\ref{tab:arch} enables a direct comparison of the three 
architectures across optical performance, thermal burden, 
operational cadence, and mass. Several conclusions emerge.

\textit{Optical performance.} Architectures~B and~C achieve 
identical optical performance --- the same 2.5~m transmitting 
aperture, coupling fraction ($f_\mathrm{coup} = 0.57$), 
energy deposited per pulse ($E_\mathrm{dep} \approx 339$~J), 
and effective velocity increment ($\Delta v \approx 0.60$~m/s)
--- despite fundamental differences in how the UV beam is
generated and delivered. Architecture~A, constrained to
1.4~m by its on-board mass budget, achieves lower coupling
($f_\mathrm{coup} = 0.25$) and correspondingly lower
$\Delta v$ (0.56~m/s), requiring either higher pulse
energy or more engagement passes per object. The 
improvement from 1.4~m to 2.5~m is therefore significant 
not only in fluence delivery but in practical mission 
efficiency.

\textit{Thermal management.} The on-board waste heat 
differs by nearly a factor of five across architectures: 
73~kW for Architecture~A (full laser on-board), 36~kW 
for Architecture~C (laser on a dedicated gondola), and 
only 15~kW for Architecture~B (THG module only). This 
directly determines the radiator sizing and the 
inter-engagement cooling interval. Architecture~A 
requires 127~s of post-engagement cooling after each 
70~s firing window --- more than twice the engagement 
duration --- which is the primary driver of its lower 
cadence. Architecture~C reduces the cooling interval 
to 7.5~s through a larger radiator allocation made 
possible by separating the laser and optics onto 
distinct gondolas. Architecture~B is essentially 
thermally unconstrained at the gondola level, since 
the 15~kW THG waste heat can be matched by a modest 
230~kg radiator.

\textit{Operational cadence and mission throughput.} 
The combined effect of engagement window and 
post-engagement cooling gives operational cadences 
of $\sim$200~s (Architecture~A), $\sim$78~s 
(Architecture~C), and $\sim$70~s under clear-sky 
conditions (Architecture~B). Over a 6-hour flight, 
this corresponds to approximately 110, 270, and up 
to 270 engagements respectively. Architecture~C 
achieves its high cadence through a combination of 
large aperture (reducing required laser power and 
thus waste heat) and a dedicated G1 gondola that 
can be fully allocated to power and thermal 
management. Architecture~B achieves comparable 
cadence but at the cost of weather dependence: 
cloud cover or strong atmospheric turbulence 
interrupts the 1030~nm uplink and reduces 
operational availability, whereas Architectures~A 
and~C are fully weather-independent.

\textit{Technology readiness and key risks.} 
The critical technology bottleneck of Architectures~A 
and~C is the compact gondola-rated kJ-class UV laser 
(TRL~2--3), which does not yet exist in a deployable 
form. Architecture~B replaces this with a 
ground-based kJ-class IR DPSSL (TRL~4--5), a 
substantially more mature technology along the 
DiPOLE or BIVOJ scaling roadmap \cite{mason2011, 
lorbeer2018}, at the cost of introducing the 
atmospheric uplink as an additional system element. 
Architecture~C additionally requires a functional 
inter-balloon optical beam transfer (TRL~4--5) 
and the mechanical operation of a 2.5~m pan-and-tilt 
on a dedicated gondola.

\textit{Summary.} No single architecture dominates 
across all metrics. Architecture~A offers the 
simplest operational concept (single gondola, 
no ground laser, no inter-balloon link) at the 
cost of the most demanding technology development 
and the lowest cadence. Architecture~B provides 
the highest near-term technology readiness and 
equivalent optical performance to Architecture~C, 
but introduces a fundamental operational 
vulnerability to weather. Architecture~C achieves 
the best overall throughput in a weather-independent 
mode, at the cost of operational complexity and 
the same laser TRL challenge as Architecture~A.

\subsection{Cost-Per-Object Estimate}

To provide an order-of-magnitude economic perspective 
we adopt the simple cost model:

\begin{equation}
C_\mathrm{obj} \approx \frac{C_\mathrm{CAPEX} + 
C_\mathrm{OPEX}}{N_\mathrm{rem}},
\label{eq:cost}
\end{equation}

where $C_\mathrm{CAPEX}$ is the total capital cost, 
$C_\mathrm{OPEX}$ the cumulative operating cost, 
and $N_\mathrm{rem}$ the total number of objects 
removed. All figures are order-of-magnitude estimates; 
detailed cost modelling is outside the scope of this 
work. We assume throughout that the required laser
technology can be engineered at the specified
performance, and that 1.5 engagement passes per
object suffice on average for deorbit of representative
thin debris fragments ($\mu \approx 1$--$5$~kg/m$^2$).
This pass count is a placeholder pending full
orbit propagation: the number of passes required
scales inversely with the effective per-pass
$\Delta v$ derived in Section~4.4, so the absolute
throughput and cost figures below should be read
as order-of-magnitude estimates. The relative
ordering across architectures is robust, since
they share the same $\Delta v$ model and differ in
per-pass increment by less than 10\%
($0.56$~m/s for~A versus $0.60$~m/s for~B and~C).

\textit{Capital costs.}
The dominant CAPEX driver for Architectures~A and~C 
is the compact gondola-rated kJ-class UV laser. 
No direct cost reference exists for such a system; 
by analogy, the DiPOLE100/BIVOJ programme (100~J, 
10~Hz) represented an investment of 
$\sim$20--50~M\$ \cite{lorbeer2018}, and the 
conceptual kJ-class DiPOLE design \cite{mason2011} 
represents a further substantial scaling step. 
For three gondola-rated 1~kJ UV systems we estimate 
$C_\mathrm{CAPEX}^\mathrm{A} \sim 400$~M\$ and 
$C_\mathrm{CAPEX}^\mathrm{C} \sim 450$~M\$. 
For Architecture~B, three ground stations with 
3--4~kJ IR DPSSL systems at 10~Hz are estimated 
at $\sim$80--150~M\$ each, giving 
$C_\mathrm{CAPEX}^\mathrm{B} \sim 300$~M\$. 
These figures are laser-dominated order-of-magnitude
estimates; the balloon platforms, telescopes, and
gondola subsystems (each well below the laser cost)
are subsumed within them. Crucially, none of the
stratospheric architectures incurs launch costs, since
all platforms are recoverable and reusable --- a
significant difference relative to orbital systems.

\textit{Operating costs.}
Annual OPEX includes balloon fabrication 
($\sim$200~k\$ per flight), launch and recovery, 
helium, staff, and maintenance, totalling 
$\sim$10--15~M\$/year for Architectures~A and~C, 
and $\sim$8--12~M\$/year for Architecture~B 
($\sim$100--150~M\$ over 10 years for each).

\textit{Mission throughput.}
Using the per-flight engagement figures from
Table~\ref{tab:arch}, 1.5 passes per object, and a
fleet-level cadence of 60 flights per year (three
reusable systems at $\sim$20 flights each, with
6-hour flights and turnaround for recovery and
refurbishment):

\begin{align}
N_\mathrm{rem}^\mathrm{A} &= 
\frac{110}{1.5} \times 60~\mathrm{fl/yr} 
\times 10~\mathrm{yr} \approx 44{,}000, \\
N_\mathrm{rem}^\mathrm{B} &= 
\frac{270}{1.5} \times 36~\mathrm{fl/yr}^* 
\times 10~\mathrm{yr} \approx 65{,}000, \\
N_\mathrm{rem}^\mathrm{C} &= 
\frac{270}{1.5} \times 60~\mathrm{fl/yr} 
\times 10~\mathrm{yr} \approx 108{,}000,
\end{align}

where $^*$60\% weather availability is assumed 
for Architecture~B.

\textit{Cost per object.}
Applying Eq.~(\ref{eq:cost}):

\begin{align}
C_\mathrm{obj}^\mathrm{A} &\approx 
\frac{400 + 125}{44{,}000} \approx 
\$12\mathrm{k}, \\
C_\mathrm{obj}^\mathrm{B} &\approx 
\frac{300 + 100}{65{,}000} \approx 
\$6\mathrm{k}, \\
C_\mathrm{obj}^\mathrm{C} &\approx 
\frac{450 + 150}{108{,}000} \approx 
\$5.5\mathrm{k}.
\end{align}

\textit{Literature comparison.}
Phipps et al.\ \cite{phipps2014} provide an explicit 
cost breakdown for L'ADROIT: total system cost 
\$560~M (\$410~M satellite, \$145~M Falcon~9 
launch, \$5~M operations over four years), 
amortised over four years. Assuming a population 
of $N_S = 100{,}000$ small debris objects and 
allocating only 10\% of system cost to the 
small-debris mission (with 90\% attributed to 
the large-debris nudging campaign for 2{,}000 
objects), the resulting cost is:

\begin{equation}
C_\mathrm{obj,small}^\mathrm{ADROIT} = 
\frac{560 \times 0.1}{100{,}000} \approx 
\$0.56\mathrm{k~per~object},
\end{equation}

giving the reported figure of ``less than \$1k'' 
\cite{phipps2014}. For large debris, 
$560 \times 0.9 / 2{,}000 \approx \$252\mathrm{k}$, 
consistent with the reported $\sim\$280\mathrm{k}$ 
\cite{phipps2014}. On a fully-loaded basis 
(all cost attributed to small debris), 
$C_\mathrm{obj}^\mathrm{ADROIT} = 
560/100{,}000 \approx \$5.6\mathrm{k}$. 
A direct quantitative comparison with the 
stratospheric estimates is complicated by 
the different cost attribution assumptions, 
campaign durations, and the orbital head-on 
engagement geometry of L'ADROIT which gives 
access rates of $\sim$19 targets/min at 
250~km range \cite{phipps2014}, substantially 
higher than the stratospheric pass geometry. 
Both figures are shown in Table~\ref{tab:econ} 
for reference.

For ground-based ablation, Scharring et al.\ 
\cite{scharring2023} show that even a 
nine-station global network operating at 
the limits of current technology falls 
significantly short of the remediation rate 
required to stabilise the LEO environment, 
making cost-per-object comparisons premature 
at this stage.

These estimates carry large uncertainties 
dominated by laser development costs and 
throughput assumptions. The relative ordering 
within the stratospheric architectures is 
nonetheless informative and is summarised in 
Table~\ref{tab:econ}.

\begin{table*}[!t]
\centering
\small
\setlength{\tabcolsep}{4pt}
\begin{tabular}{p{4.2cm}cccc}
\hline
\textbf{Parameter} & 
\textbf{A: On-board UV} & 
\textbf{B: Ground IR + UV} & 
\textbf{C: Two-balloon} & 
\textbf{L'ADROIT \cite{phipps2014}} \\
\hline
Campaign duration & 10~yr & 10~yr & 10~yr & 4~yr \\
Satellite/system CAPEX (M\$) & $\sim$400 & $\sim$300 & $\sim$450 & 410 \\
Launch cost (M\$) & 0 (recoverable) & 0 (recoverable) & 0 (recoverable) & 145 \\
Operations (M\$) & $\sim$125 & $\sim$100 & $\sim$150 & 5 \\
Total investment (M\$) & $\sim$525 & $\sim$400 & $\sim$600 & 560 \\
Target population & mixed LEO & mixed LEO & mixed LEO & 100k small + 2k large \\
$N_\mathrm{rem}$$^c$ & $\sim$44{,}000 & $\sim$65{,}000 & $\sim$108{,}000 & 100{,}000 (small) \\
$C_\mathrm{obj}$ (k\$/object)$^a$ & $\sim$12 & $\sim$6 & $\sim$5.5 & \makecell{$<$1 (10\% attr.)\\$\sim$5.6 (fully loaded)} \\
Weather dependency & None & High$^b$ & None & None \\
Platform reusability & Yes & Yes & Yes & No \\
Key TRL risk & kJ UV (2--3) & kJ IR (4--5) & kJ UV (2--3) & kJ UV (heritage) \\
\hline
\end{tabular}
\caption{Order-of-magnitude techno-economic 
comparison of the three stratospheric architectures 
and the L'ADROIT space-based reference 
\cite{phipps2014}. All figures are indicative.
$^a$L'ADROIT cost per object at 10\% cost 
attribution follows the methodology of 
\cite{phipps2014}; fully-loaded figure 
allocates total investment to small debris only.
$^b$Architecture~B operational availability
depends on clear-sky conditions for the
1030~nm ground-to-balloon uplink.
$^c$$N_\mathrm{rem}$ assumes 1.5 engagement passes
per object and a fleet-level cadence of 60 flights/yr
(36/yr effective for~B at 60\% weather availability)
over 10 years; it scales inversely with the passes
per object and is an order-of-magnitude figure.}
\label{tab:econ}
\end{table*}

\section{Conclusions}

This paper has presented a stratospheric balloon-based 
concept for laser ablation-induced momentum transfer 
to dm-scale LEO debris, and analysed three distinct 
payload architectures in terms of optical performance, 
thermal management, operational cadence, mass budget, 
and indicative cost per object removed.

Operating at $\sim$35~km altitude places the platform 
above 99\% of the atmospheric mass, substantially reducing 
turbulence, scintillation, and the ozone absorption 
that precludes UV operation from ground level. 
In this low-density propagation regime, diffraction-limited 
beam delivery at 355~nm becomes achievable with 
apertures of 1.4--2.5~m --- compatible with current 
super-pressure balloon payload capabilities --- 
and the ablation coupling efficiency is substantially 
higher than in the infrared. Unlike space-based 
systems, the platform remains fully recoverable, 
maintainable, and upgradeable between flights, 
enabling incremental technology maturation. The 
closest prior art, the relay mirror concept of 
Martin et al.\ \cite{martin2009}, retains the 
ground-based infrared laser and does not address 
UV operation from the platform; the architectures 
presented here represent a distinct step in 
which the full laser system or its UV output 
is generated at altitude.

Several constraints specific to stratospheric 
operation were identified and quantified. The
engagement range for debris at 800~km altitude
is 765~km at zenith and $\sim$830~km at the
design elevation angle $\theta^* \approx 67^\circ$
(set by the maximum slant range at which threshold
fluence is delivered, and distinct from the
$\sim$56--60$^\circ$ that maximises the impulse
efficiency), substantially larger than the 10--250~km
achievable from co-orbital platforms and 
comparable to ground-based systems, imposing 
strict requirements on laser energy and aperture. 
The balloon envelope occludes the field of view 
above $\sim$85°, reducing the usable engagement 
window to $\sim$70~s per pass. The impulse
geometry from below --- combining a retrograde
component ($\propto\cos\theta$) and a radial
component ($\propto\sin\theta$), with the radial
term contributing at one-quarter weight to perigee
lowering --- yields an effective efficiency
of $f_\mathrm{geo} \approx 0.62$ at $\theta^*$,
which is comparable to but distinct from the
head-on geometry of orbital systems.

Three payload architectures were defined and 
sized. Architecture~A (on-board 1.5~kJ UV laser,
1.4~m aperture) achieves an effective $\Delta v$ of
$\sim$0.56~m/s per pulse with an operational
cadence of $\sim$200~s per debris object, 
limited primarily by the thermal management 
of 73~kW on-board waste heat. Architecture~B
(ground 3.7~kJ IR laser with on-board UV
conversion, 2.5~m aperture) achieves an
effective $\Delta v \approx 0.60$~m/s per pulse
and a cadence
governed by atmospheric uplink availability
rather than thermal constraints, making it 
potentially the most productive architecture 
under clear-sky conditions but vulnerable 
to weather. Architecture~C (distributed 
two-balloon system, 2.5~m aperture on a 
dedicated beam director gondola) achieves 
the same $\Delta v$ as B with a thermally 
limited cadence of $\sim$78~s, yielding 
$\sim$270 engagements per 6-hour flight 
in a weather-independent mode.

The indicative techno-economic analysis 
suggests cost-per-object estimates of 
$\sim$\$12k, \$6k, and \$5.5k for 
Architectures~A, B, and C respectively 
over a 10-year campaign with a fleet of 
three systems. These estimates carry large 
uncertainties dominated by laser development 
costs. For reference, Phipps et al.\ 
\cite{phipps2014} report a cost of less 
than \$1k per small debris object for the 
L'ADROIT space-based system, based on 
allocating 10\% of the \$560M system cost 
to small debris across a population of 
100,000 objects; on a fully-loaded basis 
the equivalent figure is $\sim$\$5.6k, 
comparable to the stratospheric estimates. 
A rigorous cross-architecture comparison 
would require a unified cost model, 
consistent throughput assumptions, and 
an agreed debris population target.

The primary technology challenge common 
to Architectures~A and~C is the compact 
gondola-rated kJ-class UV laser (TRL~2--3), 
for which no deployable system currently 
exists. Architecture~B substitutes a 
ground-based kJ-class IR DPSSL 
(TRL~4--5), a more mature technology, 
at the cost of atmospheric uplink 
dependency. All three architectures 
converge on the same fundamental laser 
challenge: delivering sufficient UV 
fluence at $\sim$800~km range from 
a mass-constrained platform. Progress 
in diode-pumped solid-state laser 
technology, nonlinear frequency 
conversion efficiency, and lightweight 
thermal management --- driven by 
fusion-energy and directed-energy 
programmes --- will determine the 
practical timeline for deployment.

Within the STRATOLASER project, the 
work presented here will be followed by 
refined system-level analyses, experimental 
validation of UV propagation and ablation 
at stratospheric conditions, inter-balloon 
optical beam transfer demonstration, and 
progressive technology maturation toward 
TRL~4 through a series of stratospheric 
balloon campaigns \cite{stratolaser2025}.

\section{Acknowledgments}

This work has received funding from the European Commission under the STRATOLASER project (grant agreement No. 101223245). 

\section{Declaration of Interest}

The authors declare that a US provisional patent related to this work has been filed.

\appendix
\section{Analysis Theoretical Description}
\label{app1}

This appendix summarises the main analytical relations employed throughout the paper for
estimating spot size, aperture diameter, ablation fluence, and momentum transfer from pulsed
laser ablation.

\subsection{Diffraction-Limited Spot Size}

For a uniformly illuminated circular aperture, the diffraction-limited (Airy)
spot radius to the first null at distance $L$ is
\begin{equation}
    w_\mathrm{Airy} \approx 1.22 \, \frac{\lambda L}{D},
\end{equation}
where $\lambda$ is the wavelength and $D$ the aperture diameter. This
first-null convention is the one plotted in Figure~\ref{fig6}. The
quantitative sizing throughout this paper instead uses the
embedded-Gaussian spot radius of Eq.~(\ref{eq:spot}),
\begin{equation}
    w = \frac{M^2\lambda L}{\pi w_0} = \frac{2 M^2 \lambda L}{\pi D}
    \approx 0.64\,\frac{M^2 \lambda L}{D},
\end{equation}
i.e.\ the $1/e^2$ radius of a beam of quality $M^2$ launched from a
waist $w_0 = D/2$. For $M^2 = 1$ the two definitions differ by the
factor $1.22/(2/\pi) \approx 1.9$; all values in
Table~\ref{tab:spot} and Figure~\ref{fig:feff} use the Gaussian form.
Inverting it, the aperture required for a target spot radius $w$ is
\begin{equation}
    D \approx \frac{2 M^2 \lambda L}{\pi w}.
\end{equation}

\subsection{Strehl Ratio and Wavefront Error}

The degradation of beam quality due to wavefront aberrations is expressed using the Strehl ratio:
\begin{equation}
    S = \exp\!\left[ - \left( 2\pi \, \phi_{\mathrm{RMS}} \right)^{2} \right],
\end{equation}
where $\phi_{\mathrm{RMS}}$ is the root-mean-square wavefront error expressed in units of the
wavelength. For a given absolute aberration, the relative phase error increases as the
wavelength decreases, resulting in a lower Strehl ratio for UV compared to IR.

\subsection{Ablation Fluence Threshold}

The ablation threshold fluence $F_{\mathrm{th}}$ for nanosecond UV or IR pulses on metals
(e.g.\ aluminum) typically lies in the range:
\begin{equation}
    F_{\mathrm{th}} \sim 1{-}3 \; \mathrm{J/cm^{2}},
\end{equation}
depending on material, angle of incidence, pulse duration, and surface state. \cite{jin2013,phipps2011}
The corresponding pulse energy required at the target is:
\begin{equation}
    E_{\mathrm{p}} = F_{\mathrm{th}} \, A_{\mathrm{spot}},
\end{equation}
where $A_{\mathrm{spot}} = \pi w^{2}$ is the spot area.

\subsection{Momentum Coupling and $\Delta v$ per Pulse}

The momentum imparted by a single ablation pulse is expressed using the momentum coupling
coefficient $C_{M}$:
\begin{equation}
    M_{\mathrm{s}} \Delta v = C_{M} \, E,
\end{equation}
where $M_{\mathrm{s}}$ is the debris mass, $\Delta v$ is the velocity increment per pulse,
and $E$ is the laser energy delivered to the surface. Reported measurements of $C_M$ vary strongly with material, pulse duration, and fluence. Values in the low-fluence regimes represented in Figure~\ref{fig7} are typically around $10^{-5}$--$10^{-3}\,\mathrm{N\,s/J}$, whereas values of order $10^{-3}$--$10^{-2}\,\mathrm{N\,s/J}$ can occur near favourable fluence; we therefore use the latter only as an optimistic upper-bound sensitivity range, not as a validated baseline \cite{phipps2011,jin2013,lebras2024}.

\subsection{Total Energy Requirements}

The total energy required to produce a meaningful orbital change depends on the mission
geometry and desired $\Delta v$, but can be approximated by:
\begin{equation}
    E_{\mathrm{tot}} \approx \frac{M_{\mathrm{s}} \Delta v_{\mathrm{req}}}{C_{M}}.
\end{equation}
For representative debris masses and coupling coefficients, the resulting
energies are typically in the range of tens to hundreds of kilojoules.

\subsection{Reference Ablation Curve}

\begin{figure}[t]
\centering
\includegraphics[width=0.95\columnwidth]{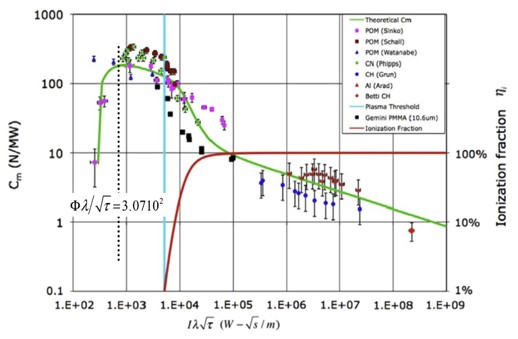}
\caption{Experimental and theoretical momentum coupling coefficients ($C_{M}$) and ionisation 
fraction as a function of laser intensity for various materials and wavelengths. The curve 
illustrates the transition from the efficient ablation regime to plasma shielding at high 
intensities, as commonly used in laser–debris interaction modelling. Data from \cite{phipps2011}.}\label{fig7}
\vspace{-2mm}
\end{figure}

\begin{figure}[t]
\centering
\includegraphics[width=0.95\columnwidth]{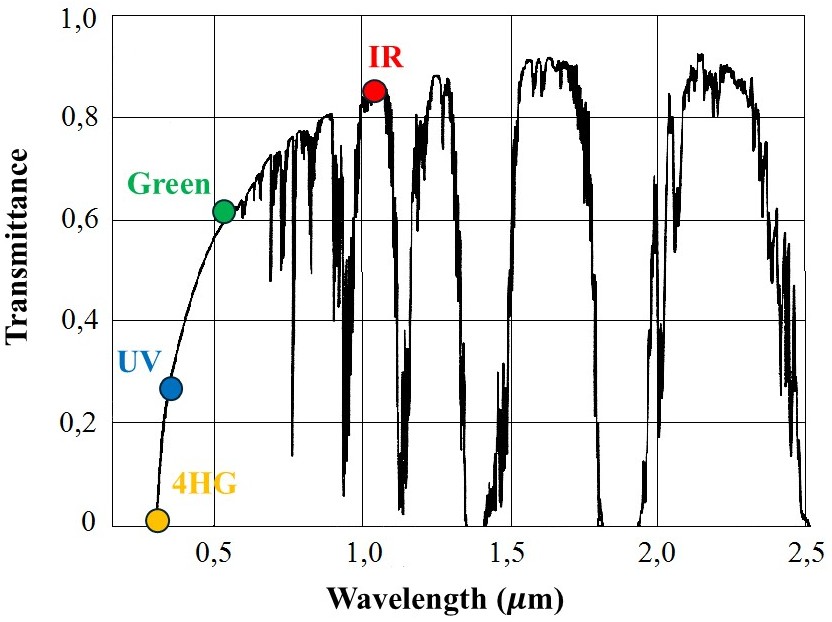}
\caption{Atmospheric transmittance as a function of wavelength at sea level. 
Short-wavelength UV and deep-UV (e.g., 355 nm and 266 nm) suffer strong Rayleigh scattering 
and molecular absorption, resulting in very low transmittance (<20\%). 
Green (532 nm) exhibits moderate transmission, while near-infrared wavelengths 
(approximately 1~\textmu m), typically used by ground-based lasers, lie near a local 
transmission maximum but still experience significant absorption bands and 
turbulence-induced distortions. }
\label{fig8}
\vspace{-2mm}
\end{figure}

Figure~\ref{fig7} summarises experimentally measured
momentum coupling coefficients and ablation regimes as a function of laser intensity, showing
the transition from efficient ablation to plasma shielding at high fluence. The corresponding sea-level spectral-transmission behaviour is shown in Figure~\ref{fig8}.

\end{document}